\documentclass[aps,prd,superscriptaddress,groupedaddress,nofootinbib,nobibnotes]{revtex4}
\usepackage{graphicx}
\usepackage{dcolumn}
\usepackage{bm}
\usepackage{amssymb}
\usepackage{epstopdf}
\usepackage{amsmath}
\usepackage{amsfonts}
\usepackage{amsthm}
\usepackage{color}
\usepackage{mathrsfs}
\usepackage{empheq}
\usepackage{dsfont}
\usepackage{float}

\usepackage{scalerel}
\usepackage{tikz}
\usetikzlibrary{svg.path}
\definecolor{orcidlogocol}{HTML}{A6CE39}
\tikzset{
  orcidlogo/.pic={
    \fill[orcidlogocol] svg{M256,128c0,70.7-57.3,128-128,128C57.3,256,0,198.7,0,128C0,57.3,57.3,0,128,0C198.7,0,256,57.3,256,128z};
    \fill[white] svg{M86.3,186.2H70.9V79.1h15.4v48.4V186.2z}
                 svg{M108.9,79.1h41.6c39.6,0,57,28.3,57,53.6c0,27.5-21.5,53.6-56.8,53.6h-41.8V79.1z M124.3,172.4h24.5c34.9,0,42.9-26.5,42.9-39.7c0-21.5-13.7-39.7-43.7-39.7h-23.7V172.4z}
                 svg{M88.7,56.8c0,5.5-4.5,10.1-10.1,10.1c-5.6,0-10.1-4.6-10.1-10.1c0-5.6,4.5-10.1,10.1-10.1C84.2,46.7,88.7,51.3,88.7,56.8z};
  }
}

\newcommand\orcidlink[1]{\href{https://orcid.org/#1}{\mbox{\scalerel*{
\begin{tikzpicture}[yscale=-1,transform shape]
\pic{orcidlogo};
\end{tikzpicture}
}{|}}}}

\newcommand\orcidicon[1]{\href{https://orcid.org/#1}{\usebox{\ORCIDlogo}}}

\usepackage[title]{appendix}
\usepackage{hyperref}
\hypersetup{colorlinks}%,allcolors=black}
\newcommand{\be}{\begin{equation}}
\newcommand{\ee}{\end{equation}}
\newcommand{\ba}{\begin{eqnarray}}
\newcommand{\ea}{\end{eqnarray}}

\newcommand{\barr}{\begin{array}}
\newcommand{\earr}{\end{array}}

\newcommand\lsim{\mathrel{\rlap{\lower4pt\hbox{\hskip1pt$\sim$}}
        \raise1pt\hbox{$<$}}}
\newcommand\gsim{\mathrel{\rlap{\lower4pt\hbox{\hskip1pt$\sim$}}
        \raise1pt\hbox{$>$}}}

\begin{document}

\title{Accurate estimation of angular power spectra for maps with correlated masks}

\author{Kristen M.~Surrao\,\orcidlink{0000-0002-7611-6179}}
\email{k.surrao@columbia.edu}
\affiliation{Department of Physics, Columbia University, New York, NY 10027, USA}

\author{Oliver~H.\,E.~Philcox\,\orcidlink{0000-0002-3033-9932}}
\affiliation{Center for Theoretical Physics, Department of Physics, Columbia University, New York, NY 10027, USA}
\affiliation{Simons Society of Fellows, Simons Foundation, New York, NY 10010, USA}

\author{J.~Colin Hill\,\orcidlink{0000-0002-9539-0835}}
\affiliation{Department of Physics, Columbia University, New York, NY 10027, USA}

\date{\today}

\begin{abstract}
The widely used MASTER approach for angular power spectrum estimation was developed as a fast $C_{\ell}$ estimator on limited regions of the sky.  This method expresses the power spectrum of a masked map (``pseudo-$C_\ell$'') in terms of the power spectrum of the unmasked map (the true $C_\ell$) and that of the mask or weight map. However, it is often the case that the map and mask are correlated in some way, such as point source masks used in cosmic microwave background (CMB) analyses, which have nonzero correlation with CMB secondary anisotropy fields and other mm-wave sky signals. In such situations, the MASTER approach gives biased results, as it assumes that the unmasked map and mask have zero correlation. While such effects have been discussed before with regard to specific physical models, here we derive a completely general formalism for any case where the map and mask are correlated. We show that our result (``reMASTERed'') reconstructs ensemble-averaged pseudo-$C_\ell$ to effectively exact precision, with significant improvements over traditional estimators for cases where the map and mask are correlated. In particular, we obtain an improvement in the mean absolute percent error from $30$\% with the MASTER result to essentially no error with the reMASTERed result for an integrated Sachs-Wolfe (ISW) field map with a mask built from the thresholded ISW field, and $10$\% to effectively zero for a Compton-$y$ map combined with an infrared source mask (the latter being directly relevant to actual data analysis). An important consequence of our result is that for maps with correlated masks, it is no longer possible to invert a simple equation to obtain the true $C_\ell$ from the pseudo-$C_\ell$. Instead, our result necessitates the use of forward modeling from theory space into the observable domain of the pseudo-$C_\ell$.  Our code is publicly available in \verb|reMASTERed|.\footnote{\url{https://github.com/kmsurrao/reMASTERed}}
\end{abstract}

\maketitle

%%%%%%%%%%%%%%%%%%%%%%%%%%%%%%%%%%%%%%%%%%%%%%%%%%%%%%%%%%%%%%%%%%
\section{Introduction}
Accurate estimation of angular power spectra is critical to determining the parameter values underlying our cosmological model. The power spectrum, or harmonic-space two-point function, of a field is perhaps the most useful summary statistic of fields considered in cosmic microwave background (CMB) analyses. Such fields include CMB secondary anisotropy fields, temperature fluctuations generated since the epoch of (baryonic) matter-radiation decoupling at redshift $z \approx 1100$ \cite{Aghanim:2007}. Examples include the thermal Sunyaev-Zel'dovich (tSZ) effect---the inverse-Compton scattering of CMB photons off hot electrons along the line of sight \cite{SZ1969, SZ1970}; the kinematic Sunyaev-Zel'dovich (kSZ) effect---the Compton scattering of CMB photons off moving electrons along the line of sight \cite{SZ1970, SZ1980, Ostriker1986}; the Integrated Sachs-Wolfe (ISW) effect---the change in CMB photon temperature due to gravitational redshifting by the late-time matter distribution \cite{Sachs:1967}; and the CMB lensing field---the deflection of CMB photons by gravitational potential wells \cite{Blanchard1987}. Moreover, power spectra are used to study other mm-wave sky signals such as the cosmic infrared background (CIB), the emission from dusty star-forming galaxies \cite{Puget1996, Gispert2000}, and radio sources, active galactic nuclei that emit radio emission.

Historically, various approaches have been considered for fast, accurate power spectrum estimation. Maximum-likelihood power spectrum estimates, although optimal in limiting regimes, depend on the data in a highly nonlinear way, requiring one to numerically solve systems of nonlinear equations and complicating likelihood analyses for the determination of cosmological parameters \cite{Tegmark:1996, Hamilton:2005}. Quadratic estimators have also been used \cite{Tegmark:1996}. With quadratic estimators, the covariance matrix that appears in the Gaussian likelihood depends linearly on the prior power spectrum, giving an analytic maximum-likelihood solution \cite{Hamilton:2005}. The difficulty with these estimators lies in their implementation, since they require knowledge of the inverse covariance between any two pixels, which becomes prohibitive for massive, high-resolution sky maps \cite{Philcox:2020,Philcox:2021ukg,PhilcoxNpoint}. 

The Monte Carlo Apodised Spherical Transform Estimator (MASTER) approach was developed to solve the need for a fast $C_{\ell}$ estimator on limited regions of the sky \cite{Hivon2002}.  The key insight of the MASTER methodology is that the power spectrum of a masked map $\tilde{T}(\mathbf{\hat{n}})$ (the so-called ``pseudo-$C_\ell$'') can be expressed in terms of the power spectrum of the unmasked map $T(\mathbf{\hat{n}})$ (the true $C_\ell$, which can be directly compared to theoretical calculations) and the power spectrum of the mask or weight map $W(\mathbf{\hat{n}})$ (denoted $C_{\ell}^{ww}$).  Letting $\sum_{\ell_2,\ell_3} \equiv \sum_{\ell_2=0}^{\infty} \sum_{\ell_3=0}^{\infty}$, the ensemble averages for the power spectra of $\tilde{T}$ and $T$ are related by
\begin{equation}
    \label{eq.MASTER_result}
    \langle \tilde{C}_{\ell_1} \rangle = \sum_{\ell_2, \ell_3} \frac{(2\ell_2+1)(2\ell_3+1)}{4\pi} C_{\ell_3}^{ww} 
    \begin{pmatrix}
    \ell_1&\ell_2&\ell_3 \\
    0&0&0
    \end{pmatrix}^2 \langle C_{\ell_2}^{aa}\rangle \,,
\end{equation}
where the penultimate factor is a Wigner $3j$ symbol, $\tilde{C}_{\ell_1}$ is the auto-power spectrum of $\tilde{T}$, $C^{aa}_{\ell_2}$ is the auto-power spectrum of $T$, and $C^{ww}_{\ell_3}$ is the auto-power spectrum of the mask $W$. Note that in the derivation of Eq.~\eqref{eq.MASTER_result} we do not require the mask to have a diagonal power spectrum; rather, we only need its diagonal components because this is enforced by the Wigner $3j$ symbols in the derivation of Eq.~\eqref{eq.MASTER_result}.

Nearly all major CMB power spectrum data analyses in the past two decades have used the MASTER result to decouple the effects of the mask from the true power spectrum (\textit{e.g.},~\cite{hinshaw2003first, Boomerang:2001, Planck:2015, Planck:2015tsz, Planck:2019, Keisler:2011, POLARBEAR:2014, Das2011, SPT-3G:2021eoc, SPT-3G:2022hvq}). Additionally, other pseudo-$C_{\ell}$ estimators (\textit{e.g.}, \cite{Elsner:2016, Tristram:2004}) and widely used software packages such as \verb|NaMaster|\footnote{\url{https://github.com/LSSTDESC/NaMaster}} \cite{Namaster} and \verb|PSpipe|\footnote{\url{https://github.com/simonsobs/PSpipe}} \cite{Li:2021} have relied on the result. A key reason for this popularity is the simple form of Eq.~\eqref{eq.MASTER_result}, which can be written as
\begin{equation}
    \label{eq.MASTER_result_simple}
    \langle \tilde{C}_{\ell_1} \rangle = \sum_{\ell_2} K_{\ell_1 \ell_2} \langle C_{\ell_2}^{aa} \rangle \,,
\end{equation}
where the mask-induced mode-coupling matrix is given by
\begin{equation}
    \label{eq.mode_coupling_matrix}
    K_{\ell_1 \ell_2} \equiv \frac{2\ell_2 + 1}{4\pi} \sum_{\ell_3} (2\ell_3+1) C_{\ell_3}^{ww} 
    \begin{pmatrix}
    \ell_1&\ell_2&\ell_3 \\
    0&0&0
    \end{pmatrix}^2 \,.
\end{equation}
Thus, after computing $K_{\ell_1 \ell_2}$ for a given sky mask, one can directly invert Eq.~\eqref{eq.MASTER_result_simple} to obtain an estimate of the true $C_\ell$ from the pseudo-$C_\ell$. This carries the assumption that the support of $K_{\ell_1 \ell_2}$ is finite, \textit{i.e.}, for the modes of interest there is little leakage from the $\ell_2$ values not sampled. With this assumption in mind, this procedure only has to be performed once, and the resulting estimate of $C_\ell^{aa}$ can then be directly used in likelihood calculations and parameter inference. 

The derivation of Eq.~\eqref{eq.MASTER_result} implicitly assumes that the map $T$ and mask $W$ are uncorrelated.  However, this assumption is not always completely valid. For example, one might want to mask out areas of a signal map that are above a certain threshold (\textit{e.g.}, \cite{Lembo:2021, wilson2012atacama}). Moreover, as is common in CMB data analyses, one may want to mask out bright radio or infrared sources, which can be correlated with the field if the field also traces large-scale structure \cite{Planck:2013lensing_CIB, Holder:2002, Shirasaki:2018, Planck:2015tsz, wilson2012atacama, Chiang:2020, Singari:2020, ACT:2015}. One frequently studied example of such a field is the tSZ effect. For example, for constructing a map of the tSZ effect from the \emph{Planck} experiment, radio and infrared point sources were masked, and the mask size was enlarged around the strongest radio sources \cite{Planck:2015tsz}. Similar masking was performed for purposes such as measuring the skewness and one-point PDF of the tSZ field from the Atacama Cosmology Telescope (ACT) \cite{wilson2012atacama, 2014arXiv1411.8004H}, studying the cosmic thermal history probed by tSZ effect tomography \cite{Chiang:2020}, and searching for the warm-hot intergalactic medium in \emph{Planck} data \cite{Singari:2020}, among several others. 

Additionally, the effects of correlated masks can potentially be highly important for cosmological fields that are sampled at the positions of galaxies. An example is cosmic shear---the distortion of images of background galaxies due to weak gravitational lensing from the large-scale structure along the line of sight (\textit{e.g.},~\cite{Kilbinger:2014cea}). Because lensing shear is only sampled at the positions of galaxies, the masks are especially complex and, moreover, highly correlated with large-scale structure due to the effects of source-lens clustering, which can become particularly significant for broad photometric redshift bins (\textit{e.g.},~\cite{Bernardeau:1997tj, nicola2021cosmic}). Similar situations can arise in fast radio burst \cite{Eftekhari:2017tbx} and peculiar velocity surveys \cite{Howlett:2022len}.

Yet another example of masking sources that are highly correlated with the field is from the analysis of the Cosmic Infrared Background Experiment (CIBER) data \cite{zemcov2014origin}. In this study, the authors were searching for near-infrared fluctuations in excess of the contributions from known galaxies; to do so, they masked such galaxies using an external catalog and deconvolved the effects of the mask using the MASTER approach, with the caveat that instead of computing the mode-coupling matrix analytically, they did so using simulations, as described in Ref.~\cite{Cooray:2012}. They then inverted the mode-coupling matrix to obtain the ``true" $C_\ell$ and ran Markov chain Monte Carlo (MCMC) methods in this space for parameter estimation. 

Previous studies have considered effects of correlated masks for tSZ effect, CIB, and radio source masks applied to the CMB lensing field, analytically modeling the correlations in real space specifically for these cases \cite{Lembo:2021, Fabbian:2020}. They predict biases on the lensed CMB temperature power spectrum with a statistical significance of well above $5\sigma$ for Simons Observatory (SO) \cite{SimonsObservatory:2018} and CMB-S4 \cite{Abazajian:2019} when masking the tSZ signal and bright regions of CIB emission. Moreover, they predict that CMB-S4 would detect the effect above $5\sigma$ for masked radio sources and that there would be potentially large biases on the polarization power spectra at subdegree scales depending on the choice of estimator.

In this paper, working in harmonic space we derive a modified version of the MASTER equation --- ``reMASTERed'' --- that includes nontrivial terms that arise when the signal map and mask are correlated. Our result is model-independent and does not require analytic simplifications needed for specific physical models. It holds exactly for \textit{any} field and mask such that the field and the portion of the mask that is correlated with the field are isotropic in the ensemble average.  For simplicity, we consider only spin-0 fields in this paper, leaving the generalization to higher-spin fields for future work.

The remainder of this paper is organized as follows.  In \S \ref{sec.notation_and_definitions}, we explain our notation and important definitions, before outlining the analytic derivation of our results in \S \ref{sec.analytic_derivation}, with details found in appendices. Next, \S \ref{sec.computational_implementation} describes the computational implementation of the result, made publicly available in  \verb|reMASTERed|. Two examples are considered in \S \ref{sec.isw_threshold} and \S \ref{sec.y_mask_ir}: an ISW effect map with a mask that masks regions with temperature above a certain threshold and a tSZ field map with an infrared source mask. Finally, \S \ref{sec.discussion} discusses our results and their implications.

%%%%%%%%%%%%%%%%%%%%%%%%%%%%%%%%%%%%%%%%%%%%%%%%%%%%%%%%%%%%%%%%%%%
\section{Notation and Definitions}
\label{sec.notation_and_definitions}

\subsection{Power Spectrum}

Denoting the map of the signal of interest (\textit{e.g.}, the CMB temperature anisotropy) as $T(\mathbf{\hat{n}})$ and the mask as $W(\mathbf{\hat{n}})$, we can write the masked map as $\tilde T(\mathbf{\hat{n}})\equiv W(\mathbf{\hat{n}})T(\mathbf{\hat{n}})$. In harmonic space, these can be expanded thus:
\begin{equation}
    \label{eq.def_TMT-tilde}
    T(\mathbf{\hat{n}}) = \sum_{\ell,m} a_{\ell m}Y_{\ell m}(\mathbf{\hat{n}}), \qquad W(\mathbf{\hat{n}}) = \sum_{\ell, m} w_{\ell m}Y_{\ell m}(\mathbf{\hat{n}}), \qquad \tilde{T}(\mathbf{\hat{n}}) = \sum_{\ell,m} \tilde{a}_{\ell m}Y_{\ell m}(\mathbf{\hat{n}}),
\end{equation}
where $\sum_{\ell, m} \equiv \sum_{\ell=0}^{\infty} \sum_{m=-\ell}^{\ell}$ and $Y_{\ell m}(\mathbf{\hat{n}})$ are spherical harmonics.

For isotropic maps and masks, the angular auto- and cross-power spectra of the above fields can be defined as
\begin{align}
    \langle a_{\ell_1 m_1}a_{\ell_2m_2}\rangle &= (-1)^{m_1} \langle a^*_{\ell_1 -m_1}a_{\ell_2m_2}\rangle \equiv (-1)^{m_1} C_{\ell_1}^{aa} \delta^{\rm K}_{\ell_1,\ell_2} \delta^{\rm K}_{m_1,-m_2} \nonumber \\
    \langle a_{\ell_1 m_1}w_{\ell_2m_2}\rangle &= (-1)^{m_1} \langle a^*_{\ell_1 -m_1}w_{\ell_2m_2}\rangle \equiv (-1)^{m_1} C_{\ell_1}^{aw} \delta^{\rm K}_{\ell_1,\ell_2} \delta^{\rm K}_{m_1,-m_2} \nonumber \\
    \langle w_{\ell_1 m_1}w_{\ell_2m_2}\rangle &= (-1)^{m_1} \langle w^*_{\ell_1 -m_1}w_{\ell_2m_2}\rangle \equiv (-1)^{m_1} C_{\ell_1}^{ww} \delta^{\rm K}_{\ell_1,\ell_2} \delta^{\rm K}_{m_1,-m_2},
\end{align} 
where $\delta^{\rm K}$ is the Kronecker delta-function.

\subsection{Bispectrum}

The connected three-point function consisting of two factors of the map and one factor of the mask is defined as \cite{Komatsu:2001rj, Fergusson:2008, fergusson2011optimal, bucher2016binned}
\begin{equation}
    \langle a_{\ell_1 m_1} a_{\ell_2 m_2} w_{\ell_3 m_3} \rangle_c \equiv  B^{\ell_1 \ell_2 \ell_3}_{m_1 m_2 m_3}[aaw] \equiv \mathcal{G}^{\ell_1 \ell_2 \ell_3}_{m_1 m_2 m_3} b_{\ell_1 \ell_2 \ell_3}^{aaw},
    \label{eq.reduced_bispectrum}
\end{equation} where $B^{\ell_1 \ell_2 \ell_3}_{m_1 m_2 m_3}[aaw]$ is the connected bispectrum, and we define the reduced bispectrum $b^{aaw}_{\ell_1 \ell_2 \ell_3}$ in the second equation. The reduced bispectrum of two different fields $a$ and $w$ is symmetric under any permutation of the joint set $\{(\ell_1,a), (\ell_2,a), (\ell_3,w)\}$.  In Eq.~\eqref{eq.reduced_bispectrum}, $\mathcal{G}^{\ell_1 \ell_2 \ell_3}_{m_1 m_2 m_3}$ is the Gaunt integral, which can be expressed in terms of Wigner $3j$ symbols as follows:
\begin{equation}
    \mathcal{G}^{\ell_1 \ell_2 \ell_3}_{m_1 m_2 m_3} \equiv \int d\mathbf{\hat{n}} \, 
    Y_{\ell_1 m_1}(\mathbf{\hat{n}}) Y_{\ell_2 m_2}(\mathbf{\hat{n}}) Y_{\ell_3 m_3}(\mathbf{\hat{n}}) = \sqrt{\frac{(2\ell_1+1)(2\ell_2+1)(2\ell_3+1)}{4\pi}} \begin{pmatrix} \ell_1&\ell_2&\ell_3 \\ 0&0&0 \end{pmatrix} \begin{pmatrix} \ell_1&\ell_2&\ell_3 \\ m_1&m_2&m_3 \end{pmatrix}
    \label{eq.def_gaunt}
\end{equation}
The Gaunt integral is symmetric under exchange of $(\ell,m)$ pairs. 

\subsection{Trispectrum}

The connected four-point function consisting of two factors of the map and two factors of the mask is defined via the reduced trispectrum as \cite{Regan:2010, Fergusson:2010}
\begin{equation}
    \label{eq.trispectrum_def}
    \langle a_{\ell_1m_1} a_{\ell_2m_2} w_{\ell_3m_3} w_{\ell_4m_4} \rangle_c \equiv \sum_{L=0}^{\infty} \sum_{M=-L}^{L} (-1)^M \mathcal{G}^{\ell_1 \ell_2 L}_{m_1 m_2 -M} \mathcal{G}^{\ell_3 \ell_4 L}_{m_3 m_4 M} t[aaww]^{\ell_1 \ell_2}_{\ell_3 \ell_4}(L) + \text{ 23 perms.},
\end{equation}
 where $t[aaww]^{\ell_1 \ell_2}_{\ell_3 \ell_4}(L)$ is the reduced (parity-even) trispectrum and the 23 permutations are taken over the joint set $\{(\ell_1,m_1,a), (\ell_2,m_2,a), (\ell_3,m_3,w), (\ell_4,m_4,w) \}$. This result is more complex than that of the bispectrum because the bispectrum is fully defined by its sides ($\ell_1,\ell_2,\ell_3$), but the trispectrum is not (assuming isotropic and homogeneous fields): one also needs the diagonal $L$. Note the following symmetries of the reduced trispectrum for two different fields $a$ and $w$
 \begin{equation}
    \label{eq.reduced_trispectrum_symmetries}
     t[aaww]^{\ell_1 \ell_2}_{\ell_3 \ell_4}(L) = t[aaww]^{\ell_2 \ell_1}_{\ell_3 \ell_4}(L) = t[wwaa]^{\ell_3 \ell_4}_{\ell_1 \ell_2}(L) \,.
 \end{equation}
 As such, there are eight permutations of the reduced trispectrum that are equal.

 We also define the trispectrum estimator 
 \begin{equation}
    \label{eq.rho_def}
     \hat{\rho}[awaw]^{\ell_1 \ell_3}_{\ell_2 \ell_4}(L) \equiv \sum_{m_1,m_2,m_3,m_4,M} (-1)^M \mathcal{G}_{m_1 m_3 -M}^{\ell_1 \ell_3 L} \mathcal{G}_{m_2 m_4 M}^{\ell_2 \ell_4 L} a_{\ell_1 m_1} a_{\ell_2 m_2} w_{\ell_3 m_3} w_{\ell_4 m_4} \,,
 \end{equation}
 which factorizes into two pieces connected by $(L,M)$. This allows for efficient implementation on data. Using Eq.~\eqref{eq.trispectrum_def}, the expectation of $\hat{\rho}[awaw]^{\ell_1 \ell_3}_{\ell_2 \ell_4}(L)$ is
 \begin{eqnarray}
     \label{eq.rho_expectation}
     \langle \hat{\rho}[awaw]^{\ell_1 \ell_3}_{\ell_2 \ell_4}(L) \rangle &=& \sum_{m_1 m_2 m_3 m_4} \sum_{L' M M'} (-1)^{M+M'} \mathcal{G}_{m_1 m_3 -M}^{\ell_1 \ell_3 L} \mathcal{G}_{m_2 m_4 M}^{\ell_2 \ell_4 L} \\\nonumber
     &&\,\times\,\left[\mathcal{G}_{m_1 m_2 -M'}^{\ell_1 \ell_2 L'} \mathcal{G}_{m_3 m_4 M'}^{\ell_3 \ell_4 L'} t[aaww]^{\ell_1 \ell_2}_{\ell_3 \ell_4}(L') + \text{ 23 perms.} \right],
 \end{eqnarray}
 where the 23 additional permutations are taken over only the piece in brackets. This estimator is the first step to measuring the trispectrum from data and is described in detail in Ref.~\citep{PhilcoxNpoint}. From the expectation of $\hat\rho$, it is clear that different $L$ and $L'$ modes are correlated, \textit{i.e.}, $\langle{\hat\rho}^{\ell_1\ell_2}_{\ell_3\ell_4}(L)\rangle$ depends on $t^{\ell_1\ell_2}_{\ell_3\ell_4}(L')$ with $L\neq L'$.\footnote{Physically, this occurs since there are two different ways to parameterize a quadrilateral using four sides and one diagonal; this is due to the two possible internal legs.} As such, the full estimator must be normalized by a deconvolution matrix which removes this correlation and additionally normalizes the estimator, such that $\langle\hat{\rho}\rangle=t$. In our application, the \textit{unnormalized} trispectrum appears in the defining reMASTERed equation; thus we omit further discussion of the deconvolution factor.

%%%%%%%%%%%%%%%%%%%%%%%%%%%%%%%%%%%%%%%%%%%%%%%%%%%%%%%%%%%%%%%%%%%
\section{ReMASTERed: Analytic Derivation}
\label{sec.analytic_derivation}

\subsection{MASTER Derivation}
Our goal is to find an expression for the power spectrum of the masked map, $\tilde{C}_\ell$, in terms of correlators of the unmasked field and the mask. As such, we start by writing out the explicit form of the expression for $\tilde{a}_{\ell m}$ (the spherical harmonic coefficients of the masked map), in terms of the map and mask:
\begin{align}
    \tilde{a}_{\ell m} &\equiv \int  d\mathbf{\hat{n}}\, T(\mathbf{\hat{n}})W(\mathbf{\hat{n}})Y^{*}_{\ell m}(\mathbf{\hat{n}})  
    = \sum_{\ell'm'}a_{\ell'm'}\int d\mathbf{\hat{n}}\, Y_{\ell'm'}(\mathbf{\hat{n}})W(\mathbf{\hat{n}})Y^{*}_{\ell m}(\mathbf{\hat{n}}) \equiv \sum_{\ell'm'}a_{\ell'm'}K_{\ell m \ell' m'} [W],
\end{align}
where $K_{\ell m \ell' m'} [W]$ is the mode-coupling kernel with\footnote{This definition corrects a typo in Appendix A.2 of Ref.~\cite{Hivon2002}.}

\begin{align}
    K_{\ell_1 m_1 \ell_2 m_2} [W] &\equiv \int d\mathbf{\hat{n}}\, Y^*_{\ell_1 m_1}(\mathbf{\hat{n}})W(\mathbf{\hat{n}})Y_{\ell_2 m_2}(\mathbf{\hat{n}})
    =\sum_{\ell_3m_3}w_{\ell_3m_3}(-1)^{m_1} \mathcal{G}^{\ell_1 \ell_2 \ell_3}_{-m_1 m_2 m_3},
\end{align}
expanding the mask in spherical harmonics. Using our expression for $\tilde{a}_{\ell m}$, we can compute the auto-spectrum of the masked map
\begin{align}
    \langle \tilde{C}_{\ell_1} \rangle &\equiv \frac{1}{2\ell_1+1}\sum_{m_1=-\ell_1}^{\ell_1} \langle \tilde{a}_{\ell_1m_1} \tilde{a}^{*}_{\ell_1m_1} \rangle %\nonumber
    %\\&=
    =\frac{1}{2\ell_1+1}\sum_{m_1=-\ell_1}^{\ell_1}\sum_{\ell_2 m_2}\sum_{\ell_3 m_3} \langle K_{\ell_1 m_1 \ell_2 m_2}[W] K^{*}_{\ell_1 m_1 \ell_3 m_3}[W]  a_{\ell_2 m_2}a^{*}_{\ell_3 m_3} \rangle 
\end{align}
The MASTER approach then assumes that the signal map and mask are not correlated so that the RHS of this equation can be separated as follows~\cite{Hivon2002}:
\begin{align}
    \langle \tilde{C}_{\ell_1} \rangle &= \frac{1}{2\ell_1+1}\sum_{m_1=-\ell_1}^{\ell_1}\sum_{\ell_2 m_2}\sum_{\ell_3 m_3} \langle K_{\ell_1 m_1 \ell_2 m_2}[W] K^{*}_{\ell_1 m_1 \ell_3 m_3}[W] \rangle  \langle a_{\ell_2 m_2}a^{*}_{\ell_3 m_3} \rangle \nonumber \\
    &= \frac{1}{2\ell_1+1}\sum_{m_1=-\ell_1}^{\ell_1}\sum_{\ell_2 m_2}\sum_{\ell_3 m_3} \sum_{\ell_4 m_4} \sum_{\ell_5 m_5} \langle a_{\ell_2 m_2}a^{*}_{\ell_3 m_3} \rangle \langle w_{\ell_4 m_4} w^{*}_{\ell_5 m_5} \rangle \mathcal{G}^{\ell_1 \ell_2 \ell_4}_{-m_1 m_2 m_4} \mathcal{G}^{\ell_1 \ell_3 \ell_5}_{-m_1 m_3 m_5} 
\end{align} and proceeds to obtain the result in Eq.~\eqref{eq.MASTER_result}. 

\subsection{ReMASTERed Corrections to MASTER Derivation}
However, if we allow for the possibility that the signal map and mask may be correlated, the above separation is not possible, and we must evaluate the full four-point function involving two factors of the map and two factors of the mask. To make contact with the original MASTER result and to gain intuition for what is happening, we explicitly write out the Wick contractions of the four-point function to obtain terms such as the cross-spectra of the signal and mask, as well as connected bispectrum and trispectrum terms:
\begin{align}
    \langle \tilde{C}_{\ell_1} \rangle &= \frac{1}{2\ell_1+1}\sum_{m_1=-\ell_1}^{\ell_1}\sum_{\ell_2 m_2}\sum_{\ell_3 m_3} \sum_{\ell_4 m_4} \sum_{\ell_5 m_5} \langle a_{\ell_2 m_2}a^{*}_{\ell_3 m_3} w_{\ell_4 m_4} w^{*}_{\ell_5 m_5} \rangle \mathcal{G}^{\ell_1 \ell_2 \ell_4}_{-m_1 m_2 m_4} \mathcal{G}^{\ell_1 \ell_3 \ell_5}_{m_1 -m_3 -m_5} \nonumber
    \\&= \frac{1}{2\ell_1+1} \sum_{m_1=-\ell_1}^{\ell_1}\sum_{\ell_2 m_2}\sum_{\ell_3 m_3} \sum_{\ell_4 m_4} \sum_{\ell_5 m_5} (-1)^{m_1} \mathcal{G}^{\ell_1 \ell_2 \ell_4}_{-m_1 m_2 m_4} \mathcal{G}^{\ell_1 \ell_3 \ell_5}_{m_1 m_3 m_5} 
    [ \langle a_{\ell_2 m_2}a_{\ell_3m_3}\rangle  \langle w_{\ell_4 m_4}w_{\ell_5m_5}\rangle \notag\\ &\qquad + \langle a_{\ell_2 m_2}w_{\ell_4m_4} \rangle \langle a_{\ell_3 m_3}w_{\ell_5m_5} \rangle + \langle a_{\ell_2 m_2}w_{\ell_5m_5} \rangle \langle a_{\ell_3 m_3}w_{\ell_4m_4} \rangle + \langle w_{\ell_4 m_4} \rangle \langle a_{\ell_2 m_2} a_{\ell_3 m_3} w_{\ell_5 m_5} \rangle_c \nonumber \notag\\ &\qquad + \langle w_{\ell_5 m_5} \rangle \langle a_{\ell_2 m_2} a_{\ell_3 m_3} w_{\ell_4 m_4} \rangle_c + \langle a_{\ell_2 m_2} \rangle \langle a_{\ell_3 m_3} w_{\ell_4 m_4} w_{\ell_5 m_5} \rangle_c + \langle a_{\ell_3 m_3} \rangle \langle a_{\ell_2 m_2} w_{\ell_4 m_4} w_{\ell_5 m_5}\rangle_c \nonumber \notag\\ &\qquad  + \langle a_{\ell_2 m_2} a_{\ell_3 m_3} w_{\ell_4 m_4} w_{\ell_5 m_5} \rangle_c] \label{eq.split_into_npoint}.
\end{align}
We have not made any assumptions about $\bar{a}$ or $\bar{w}$, the map and mask averages, respectively.

First, consider the trispectrum term (the final term in Eq.~\eqref{eq.split_into_npoint}). Using the definition in Eq.~\eqref{eq.trispectrum_def}, this term becomes
\begin{equation}
    \frac{1}{2\ell_1+1} \sum_{m_1=-\ell_1}^{\ell_1}\sum_{\ell_2 m_2}\sum_{\ell_3 m_3} \sum_{\ell_4 m_4} \sum_{\ell_5 m_5} \sum_{LM} (-1)^{m_1+M} \mathcal{G}^{\ell_1 \ell_2 \ell_4}_{-m_1 m_2 m_4} \mathcal{G}^{\ell_1 \ell_3 \ell_5}_{m_1 m_3 m_5} \left[\mathcal{G}^{\ell_2 \ell_3 L}_{m_2 m_3 -M} \mathcal{G}^{\ell_4 \ell_5 L}_{m_4 m_5 M} t[aaww]^{\ell_2 \ell_3}_{\ell_4 \ell_5}(L) + \text{ 23 perms.}\right]
\end{equation}
Comparing to Eq.~\eqref{eq.rho_expectation}, we see that this term is just
\begin{equation}
    \frac{1}{2\ell_1+1} \sum_{\ell_2 \ell_3 \ell_4 \ell_5} \langle \hat{\rho}[awaw]^{\ell_2 \ell_4}_{\ell_3 \ell_5}(\ell_1) \rangle \,.
\end{equation}
This is the unnormalized trispectrum estimator, which can be measured directly from the data or simulations.

Using orthogonality relations and properties of the Gaunt integrals, the entire equation for $\langle \tilde{C}_{\ell_1} \rangle$ simplifies to 

\begin{equation}
\boxed{
\begin{aligned}
     \langle \tilde{C}_{\ell_1} \rangle 
    &= \frac{1}{4\pi} \sum_{\ell_2,\ell_3} (2\ell_2+1)(2\ell_3+1) \begin{pmatrix} \ell_1&\ell_2&\ell_3 \\ 0&0&0 \end{pmatrix}^2 \left[  \langle C_{\ell_2}^{aa}\rangle
    \langle C_{\ell_3}^{ww}\rangle +  \langle C_{\ell_2}^{aw}\rangle  \langle C_{\ell_3}^{aw}\rangle + \frac{\langle w_{00} \rangle}{\sqrt{\pi}}  \langle b_{\ell_1 \ell_2 \ell_3}^{aaw} \rangle + \frac{\langle a_{00} \rangle}{\sqrt{\pi}}  \langle b_{\ell_1 \ell_2 \ell_3}^{waw} \rangle  \right]  \\&\qquad + \frac{1}{2\ell_1+1} \sum_{\ell_2 \ell_3 \ell_4 \ell_5} \langle \hat{\rho}[awaw]^{\ell_2 \ell_4}_{\ell_3 \ell_5}(\ell_1) \rangle, 
     \label{eq.full_result_with_rho}
\end{aligned}
}
\end{equation}
where $\langle \hat{\rho}[awaw]^{\ell_2 \ell_4}_{\ell_3 \ell_5}(\ell_1) \rangle$ is the ensemble-averaged expectation of $\hat{\rho}$.  This equation is derived in full in Appendix \ref{sec.appendix_full_derivation_iso} and is the main result of this work. Schematically, we refer to each of the terms on the RHS of Eq.~\eqref{eq.full_result_with_rho} as the $\langle aa \rangle \langle ww \rangle$, $\langle aw \rangle \langle aw \rangle$, $\langle w \rangle \langle aaw \rangle_c$, $\langle a \rangle \langle waw \rangle_c$, and $\langle aaww \rangle_c$ terms, respectively.  The first term alone corresponds to the standard MASTER result in Eq.~\eqref{eq.MASTER_result}, while all of the others are generated by non-zero correlations between the unmasked field and the mask.

In the above, we have written the trispectrum term in terms of the $\hat\rho$ quantity that would be estimated from simulations. In practical settings it may be more useful to write this in terms of the theoretical reduced trispectrum, expanding 
$\langle \hat{\rho}[awaw]^{\ell_2 \ell_4}_{\ell_3 \ell_5}(\ell_1) \rangle$. This yields
\begin{align}
    \label{eq.full_result_with_trispectrum}
    \langle \tilde{C}_{\ell_1} \rangle 
    &= \frac{1}{4\pi} \sum_{\ell_2,\ell_3} (2\ell_2+1)(2\ell_3+1) \begin{pmatrix} \ell_1&\ell_2&\ell_3 \\ 0&0&0 \end{pmatrix}^2 \left[  \langle C_{\ell_2}^{aa}\rangle
    \langle C_{\ell_3}^{ww}\rangle +  \langle C_{\ell_2}^{aw}\rangle  \langle C_{\ell_3}^{aw}\rangle + \frac{\langle w_{00} \rangle}{\sqrt{\pi}}  \langle b_{\ell_1 \ell_2 \ell_3}^{aaw} \rangle + \frac{\langle a_{00} \rangle}{\sqrt{\pi}}  \langle b_{\ell_1 \ell_2 \ell_3}^{waw} \rangle  \right] \\\notag \nonumber &+ \frac{8}{(4\pi)^2} \sum_{\ell_2 \ell_3 \ell_4 \ell_5} (2\ell_2+1)(2\ell_3+1)(2\ell_4+1)(2\ell_5+1) \begin{pmatrix} \ell_1&\ell_2&\ell_4 \\ 0&0&0 \end{pmatrix} \begin{pmatrix} \ell_1&\ell_3&\ell_5 \\ 0&0&0 \end{pmatrix} \notag \nonumber \\&\times \Bigg[ \sum_L (-1)^{\ell_1+L}  \begin{Bmatrix} \ell_4&\ell_2&\ell_1 \\ \ell_3&\ell_5&L \end{Bmatrix} \begin{pmatrix} \ell_2&\ell_3&L \\ 0&0&0 \end{pmatrix} \begin{pmatrix} \ell_4&\ell_5&L \\ 0&0&0 \end{pmatrix} t[aaww]^{\ell_2 \ell_3}_{\ell_4 \ell_5}(L)  \notag \nonumber \\&\;\;\; + \sum_L (-1)^{\ell_1+L}  \begin{Bmatrix} \ell_4&\ell_2&\ell_1 \\ \ell_5&\ell_3&L \end{Bmatrix} \begin{pmatrix} \ell_2&\ell_5&L \\ 0&0&0 \end{pmatrix} \begin{pmatrix} \ell_3&\ell_4&L \\ 0&0&0 \end{pmatrix} t[awaw]^{\ell_2 \ell_5}_{\ell_3 \ell_4}(L) \notag \nonumber \\&\;\;\; + \begin{pmatrix} \ell_1&\ell_2&\ell_4 \\ 0&0&0 \end{pmatrix} \begin{pmatrix} \ell_1&\ell_3&\ell_5 \\ 0&0&0 \end{pmatrix} t[awaw]^{\ell_3 \ell_5}_{\ell_2 \ell_4}(\ell_1) \Bigg]\nonumber, 
\end{align}
where $\begin{Bmatrix} \ell_4&\ell_2&\ell_1 \\ \ell_3&\ell_5&L \end{Bmatrix}$ is a Wigner $6j$ symbol. If the map is a zero-mean field, the term involving $\langle a_{00} \rangle$ goes to zero in both Eq.~\eqref{eq.full_result_with_rho} and Eq.~\eqref{eq.full_result_with_trispectrum}.

\subsection{Comments on ReMASTERed Result}
An important consideration for the above results concerns which factors require the maps and masks to have their ensemble averages subtracted. To understand this further, we consider the following schematic argument.  Denote some map as $X=\langle X \rangle + \delta X$, where $\langle{X\rangle}$ is the ensemble averaged expectation value.\footnote{Note that we express a map in terms of its ensemble average and deviation instead of the individual map's mean and deviation. If we instead use the mean, we get terms like $\langle \bar{w} \delta a \rangle$, where $\bar{w}$ is the mean of the mask. Using ensemble averages is cleaner since such terms separate as $\langle w \rangle \langle \delta a \rangle = 0$.} Thus
\begin{equation}
    \tilde{a} =  aw = \left(\langle a \rangle + \delta a \right) \left(\langle w \rangle + \delta w \right) 
    \label{eq.schematic_atilde}
\end{equation}
and 
\begin{align}
    \langle \tilde{C}_{\ell_1} \rangle  \sim \langle \tilde{a} \tilde{a} \rangle &= \underbrace{\langle a \rangle \langle a \rangle \langle w \rangle \langle w \rangle + \langle \delta a \delta a \rangle \langle w \rangle \langle w \rangle + \langle \delta w \delta w \rangle \langle a \rangle \langle a \rangle + \langle \delta a \delta a \rangle \langle \delta w \delta w \rangle}_{\langle aa \rangle \langle ww \rangle } + \underbrace{2\langle \delta a \delta w \rangle \langle \delta a \delta w \rangle + 4\langle \delta a \delta w \rangle \langle a \rangle \langle w \rangle}_{\mathclap{{} 2\langle aw \rangle \langle aw \rangle }} \nonumber
    \\&+ \underbrace{2 \langle w \rangle \langle \delta a \delta a \delta w \rangle_c + 2 \langle a \rangle \langle \delta w \delta a \delta w \rangle_c}_{\sim \mathrm{bispectrum \; terms}} + \underbrace{\langle \delta a \delta a \delta w \delta w \rangle_c}_{\sim \mathrm{trispectrum \; term}} \label{eq.schematic_where_mean_subt}
\end{align}
The second piece in Eq.~\eqref{eq.schematic_where_mean_subt} is actually $2[ \langle aw \rangle \langle aw \rangle - \langle a \rangle \langle a \rangle \langle w \rangle \langle w \rangle]$, but the latter term
only affects the result at $\ell_1=0$ and can thus be ignored. The result in Eq.~\eqref{eq.schematic_where_mean_subt} then implies that the ensemble averages $\langle a \rangle$ and $\langle w \rangle$ must be subtracted from the map and mask, respectively, of each realization only before entering the bispectrum and trispectrum calculations but not before entering the two-point functions.

We note that this derivation has assumed isotropy of the signal and correlated mask in the ensemble average. The usual MASTER result does not require such an assumption for the mask since the Wigner $3j$ symbols force only the $\ell=\ell'$ parts of $\langle w_{\ell m} w_{\ell' m'} \rangle$ to contribute. In practice, one may have some mask that is correlated with a map but also has separable anisotropic contributions. Our result still gives correct results for such a mask, \textit{i.e.}, it holds regardless of the isotropy of the overall mask, so long as the piece of the mask that is correlated with the field is isotropic. To see this, consider the following argument.

The mask can be thought of as two separate sub-masks: one that is statistically anisotropic but uncorrelated with the map (\textit{e.g.}, a Galactic mask in a CMB map), and one that is isotropic but correlated with the map, \textit{i.e.},
\begin{equation}
    \label{eq.split_mask}
    \tilde{T}(\mathbf{\hat{n}}) = T(\mathbf{\hat{n}})W(\mathbf{\hat{n}}) = T(\mathbf{\hat{n}})[W^{\rm aniso}(\mathbf{\hat{n}}) + W^{\rm iso}(\mathbf{\hat{n}}) ]
\end{equation}
Then schematically, 
\begin{align}
    \langle \tilde{C}_{\ell_1} \rangle  \sim \langle \tilde{a} \tilde{a} \rangle &= \langle aa(w^{\rm aniso} + w^{\rm iso})(w^{\rm aniso} + w^{\rm iso}) \rangle 
    \\&= \underbrace{\langle aa \rangle \left[\langle w^{\rm iso}w^{\rm iso} \rangle + w^{\rm aniso}w^{\rm aniso} + 2 \langle w^{\rm iso} \rangle w^{\rm aniso}\right]}_{\langle aa \rangle \langle ww \rangle} + \underbrace{2 \langle a w^{\rm iso} \rangle \left[\langle a w^{\rm iso} \rangle + 2 \langle a \rangle w^{\rm aniso} \right]}_{2\langle aw \rangle \langle aw \rangle} \nonumber
    \\&\; + \underbrace{2\langle aaw^{\rm iso} \rangle_c \left[\langle w^{\rm iso} \rangle + w^{\rm aniso} \right]}_{2\langle w \rangle \langle aaw \rangle_c} + \underbrace{2\langle a \rangle \langle w^{\rm iso}aw^{\rm iso} \rangle_c}_{2\langle a \rangle \langle waw \rangle_c} + \underbrace{\langle aaw^{\rm iso}w^{\rm iso} \rangle_c}_{\langle aaww \rangle_c} \nonumber
\end{align}
We consider each of the pieces in the above that involve the anisotropic, uncorrelated component of the mask and show that the anisotropy does not pose any issue for our derivation, with details in Appendix \ref{sec.appendix_full_derivation_iso}. The first piece is the $\langle aa \rangle \langle ww \rangle$ term, the usual MASTER result. As previously mentioned, the MASTER derivation does not require isotropy of the mask. The second piece is the $\langle aw \rangle \langle aw \rangle$ term.\footnote{This piece is actually $2[\langle aw \rangle \langle aw \rangle - \langle a \rangle \langle a \rangle w^{\rm aniso} w^{\rm aniso}]$, but the latter term only affects the result at $\ell_1=0$ and can thus be ignored, similar to the case with the $\langle a \rangle \langle a \rangle \langle w \rangle \langle w \rangle$ term in Eq.~\eqref{eq.schematic_where_mean_subt}.} The anisotropic term in this piece is $4\langle a w^{\rm iso} \rangle \langle a \rangle w^{\rm aniso}$. Starting from the relevant terms in Eq.~\eqref{eq.appendix_full_expansion} in Appendix \ref{sec.appendix_full_derivation_iso}, it can be shown that only the $w^{\rm aniso}_{00}$ spherical harmonic coefficient contributes to the result, without requiring isotropy of $w^{\rm aniso}$. Finally, the third piece is the $\langle w \rangle \langle aaw \rangle_c$ term. From Eq.~\eqref{eq.term4_lastline}, the Gaunt factors force only $w^{\rm aniso}_{00}$ to contribute to the result, again without requiring isotropy of $w^{\rm aniso}$. These details allow the anisotropic contributions to the mask to go through our derivation without modifying the results, as long as the anisotropic contributions are not correlated with the map. Of note is that the term involving the connected trispectrum holds regardless of the isotropy of the mask (and thus holds even for anisotropically correlated portions of the mask). To see this, instead of parameterizing the trispectrum by four $\ell$ values and four $m$ values, one can imagine parameterizing the trispectrum by four $\ell$ values and some external $L$ and $M$ coefficients, which parameterize the angular momentum symmetries of the entire expression. Averaging over rotations, as we do for obtaining $\langle \tilde{C}_{\ell_1} \rangle$, forces $L=M=0$, which is exactly the basis of the isotropic reduced trispectrum that we have used.

%%%%%%%%%%%%%%%%%%%%%%%%%%%%%%%%%%%%%%%%%%%%%%%%%%%%%%%%%%%%%%%%%%%
\section{Computational Implementation}
\label{sec.computational_implementation}
Our code \verb|reMASTERed| computes the expression on the RHS of Eq.~\eqref{eq.full_result_with_rho} for the power spectrum of the masked map.  The code has functionality both to compute the results given an arbitrary map and mask and also to compute the results in the ensemble average for the type of threshold mask described in \S \ref{sec.isw_threshold}. We choose to compute Eq.~\eqref{eq.full_result_with_rho} as opposed to Eq.~\eqref{eq.full_result_with_trispectrum} because $\langle \hat{\rho}[awaw]^{\ell_2 \ell_4}_{\ell_3 \ell_5}(\ell_1) \rangle$ can be measured directly from our simulations, and we thus avoid having to compute a nondiagonal trispectrum normalization matrix. In all practical scenarios, we would have a theoretical model to use for the reduced trispectrum; this use of $\hat{\rho}$ is just for testing purposes here.

For calculation of the bispectrum and $\hat{\rho}$, we adapt code from \verb|PolyBin|\footnote{\url{https://github.com/oliverphilcox/PolyBin}} \cite{PhilcoxNpoint} without $\ell$-space binning, as the bispectrum and $\hat{\rho}$ involving both $a$ and $w$ may exhibit large fluctuations as $\ell$ varies. This has a tradeoff: the lack of binning slows runtime and increases memory usage. For testing purposes we thus limit ourselves to low values of $\ell_{\rm max}$, defined as the maximum $\ell$ in the $\ell_2,\ell_3,\ell_4,\ell_5$ summations and calculations of $\langle \tilde{C}_{\ell} \rangle$.  Because of these sums, calculation of our results scales as $\mathcal{O}(\ell_{\rm max}^4)$. In practice, if one has smooth theoretical models for the power spectra, bispectra, and trispectra involving the different factors of the unmasked map and mask, it would be possible to bin in $\ell$ to reduce the dimensionality of the problem, allowing one to push to much higher $\ell_{\rm max}$.

The overall idea of the validation that we perform using our code is as follows: 

\begin{enumerate}
    \item Generate several independent random realizations of some initial field. For a Gaussian random field, this is implemented in \verb|HEALPix/healpy| \cite{Healpix, Healpy} given an initial power spectrum. For the tSZ effect, one can use \verb|halosky| or other packages to populate a simulated sky map with projected galaxy cluster pressure profiles.\footnote{\url{https://github.com/marcelo-alvarez/halosky}}  
    \item  Define some masking operation that is correlated with that field and apply this masking operation to each realization independently (for example, we might make a threshold mask that masks out pixels in the map realization above a certain value).\footnote{For testing purposes, we define some $\ell_{\rm max}$ for which we will calculate $\tilde{C}_{\ell}$ and then remove any power in the map and mask above $\ell_{\rm max}$, \textit{i.e.}, we work only with band-limited data. The reason for this is that, theoretically, the values over which we sum $\ell_2, \ell_3, \ell_4, \ell_5$ in Eq.~\eqref{eq.full_result_with_rho} can be arbitrarily large so long as the triangle conditions imposed by the Wigner $3j$ symbols are satisfied. Therefore, by removing power in both the map and mask beyond $\ell_{\rm max}$, we ensure that only modes below $\ell_{\rm max}$ can contribute to the final result. Note that this is just a computational trick to test our results. In practice, one can instead choose an $\ell$-range containing all values for which the relevant functions have non-negligible support.}
    \item For each realization, measure $a_{00}$, $w_{00}$, $C_{\ell}^{aa}$, $C_{\ell}^{ww}$, and $C_{\ell}^{aw}$. Further compute $b^{aaw}_{\ell_1 \ell_2 \ell_3}$, $b^{waw}_{\ell_1 \ell_2 \ell_3}$, and $\langle \hat{\rho}[awaw]^{\ell_2 \ell_4}_{\ell_3 \ell_5}(\ell_1) \rangle$ with the ensemble average subtracted from each map and mask before entering the calculation. Finally, directly compute $\tilde{C}_{\ell}$ from the masked sky map for comparison (\textit{i.e.}, calculate the power spectrum of the masked sky map).
    \item Average each of these quantities over all the realizations to obtain the ensemble averages and then insert these results into Eq.~\eqref{eq.full_result_with_rho} to obtain the final result, \textit{i.e.}, a comparison of the LHS and RHS of that equation, with the LHS labeled as ``directly computed.''  Wigner $3j$ symbols are calculated with \verb|pywigxjpf|.\footnote{\url{https://pypi.org/project/pywigxjpf/}}
\end{enumerate}
 Several consistency checks of the result are provided in Appendix \ref{sec.appendix_consistency_checks}.

%%%%%%%%%%%%%%%%%%%%%%%%%%%%%%%%%%%%%%%%%%%%%%%%%%%%%%%%%%%%%%%%%%%
\section{Application: ISW Field with Threshold Mask}
\label{sec.isw_threshold}

\begin{figure}[t]
    \centering
    \includegraphics[width=0.9\textwidth]{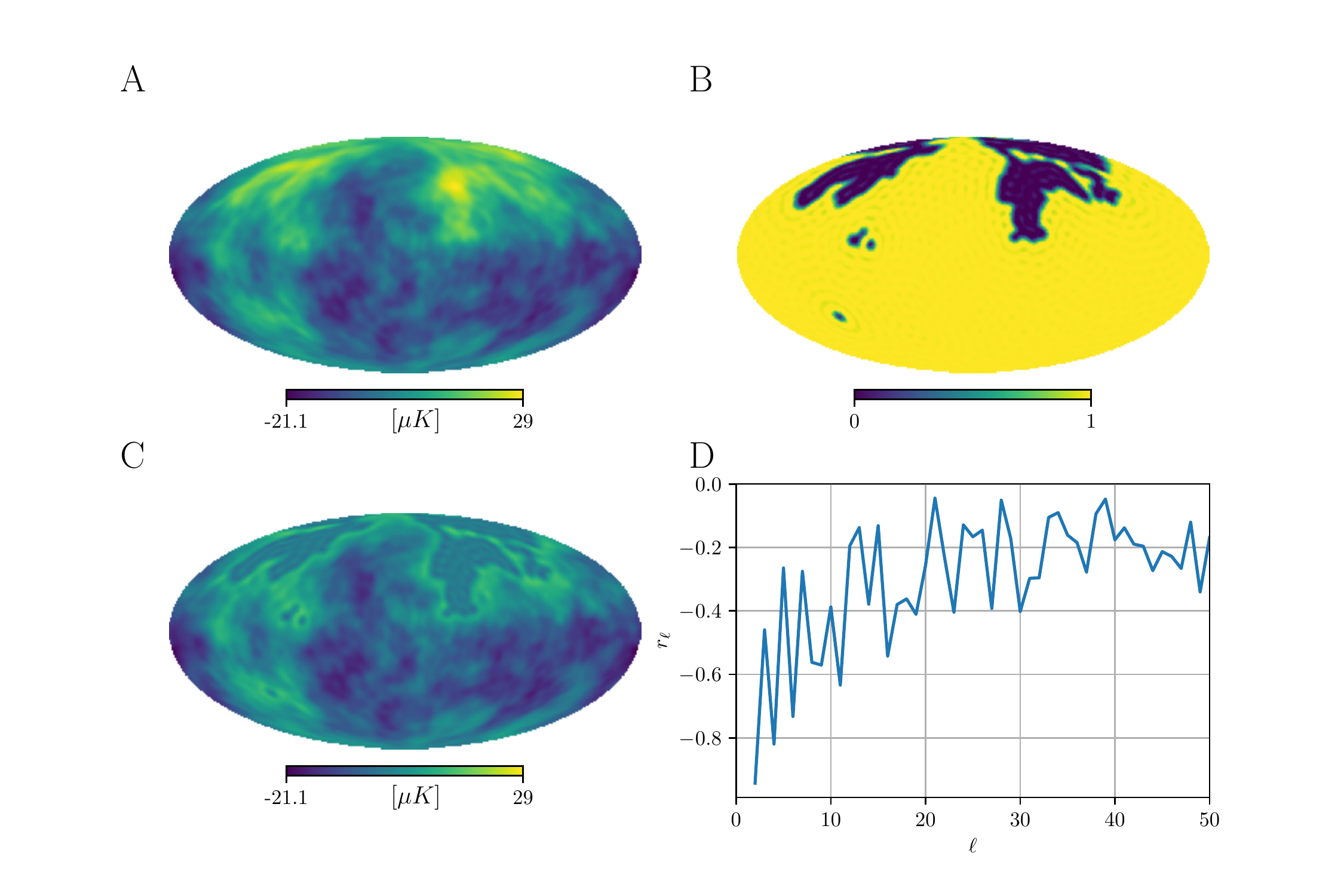}
    \caption{\textbf{A.} WebSky ISW map (in $\mu$K units) with $N_{\rm side}=128$. \textbf{B.} Mask designed to mask out regions of the ISW map with values greater than 1.5 standard deviations of the map's mean and then apodized with a cosine apodization of width $2.5^{\circ}$. \textbf{C.} Masked ISW map [$\mu$K]. \textbf{D.} Correlation coefficient of ISW map and threshold mask. We note significant correlations (which are ignored in the original MASTER approach to power spectrum estimation) at low $\ell$.}
    \label{fig:ISW_maps}
\end{figure}

The Integrated Sachs-Wolfe (ISW) effect---the change in CMB photon temperature due to gravitational redshifting by the late-time matter distribution---has been detected via the cross-correlation of CMB and large-scale structure data (see \cite{Nishizawa:2014} for a review). It has been measured via cross-correlation of the CMB with WISE \cite{Ferraro:2014},  AllWISE \cite{Shajib:2016}, and unWISE \cite{Krolewski:2021} galaxies and various other large-scale structure tracers \cite{Planck:2015isw}.  The ISW signal is predominantly found on large angular scales.  For a detection of the ISW effect, one often wants to mask bright point sources to reduce contamination in the CMB map \cite{Krolewski:2021}, and these can be highly correlated with the field \cite{Cooray:2001}. In some cases, this may amount to thresholding the field by masking out bright areas of the field. We use this thresholding procedure for demonstration purposes here, due to its simplicity.

For our implementation, an ISW field map is obtained from the WebSky Extragalactic CMB Mocks\footnote{\url{https://mocks.cita.utoronto.ca/index.php/WebSky_Extragalactic_CMB_Mocks}} \cite{Websky2020}. The WebSky simulation suite combined the mass-Peak Patch approach \cite{Stein:2018lrh,Bond1996} with halo occupation distribution (HOD) models for various observable quantities to ``paint'' the components on the halos. The authors used a realization of the cosmic web for redshifts $0 < z < 4.6$ over the full sky and a volume of $\sim 600 \, (\mathrm{Gpc}/h)^3$ with $\sim 10^{12}$ resolution elements \cite{Websky2020}. To reduce computational expense, we downgrade the ISW map to $N_{\rm side} = 128$. A mask is generated by initially setting the mask value at each pixel to 1. Then, pixels corresponding to regions where the ISW signal is above $1.5$ standard deviations of the map's mean are set to 0. To reduce sensitivity to individual pixels with outlier values, a temporary map---equal to the original map but downgraded to $N_{\rm side}=64$---is created to identify regions to mask. The mask is apodized using \verb|NaMaster| \cite{Namaster} with a cosine apodization of width $2.5^{\circ}$. The mask is then upgraded to $N_{\rm side}=128$ to match the resolution of the original map. Figure \ref{fig:ISW_maps} shows the ISW field map and the corresponding mask. The correlation coefficient $r_{\ell}$ of the map and mask, calculated as $r_{\ell} \equiv C_{\ell}^{aw}/\sqrt{C_{\ell}^{aa}C_{\ell}^{ww}}$, is also shown in Figure \ref{fig:ISW_maps}.

\begin{figure}[htb]
    \centering
    \includegraphics[width=0.80\textwidth]{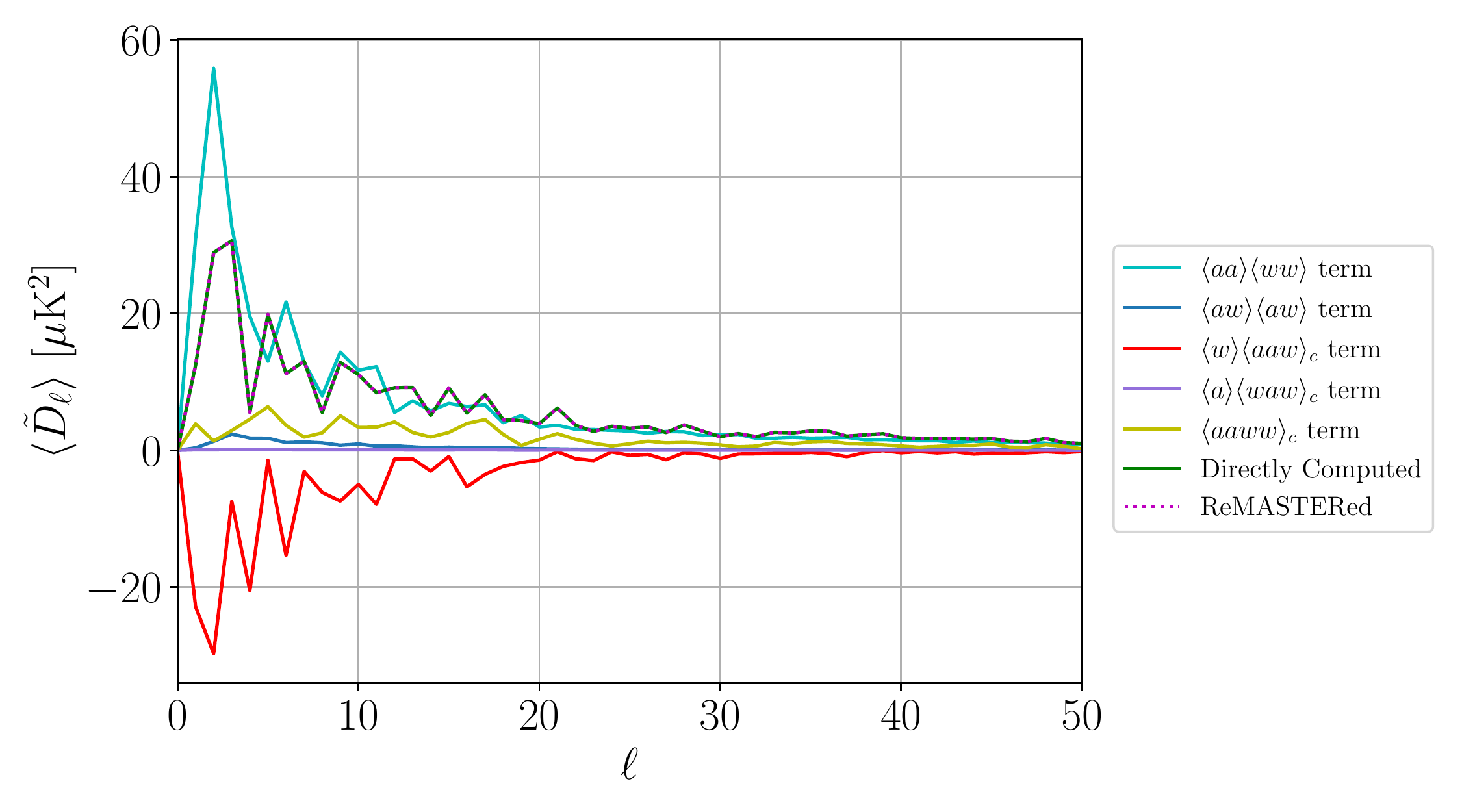}
    \caption{Comparison of MASTER and reMASTERed pseudo-$C_\ell$ reconstruction (implementation of Eq.~\eqref{eq.full_result_with_rho}) for the ISW field with threshold mask. The directly computed power spectrum of the masked map (solid green) is compared to our reMASTERed result (dotted magenta) as a demonstration of the validity of our result and the accuracy of our computational implementation. The relative contributions of each of the terms in the reMASTERed result are also shown: $\langle aa \rangle \langle ww \rangle$ term (solid cyan, the original MASTER result), $\langle aw \rangle \langle aw \rangle$ term (solid blue), $\langle w \rangle \langle aaw \rangle_c$ term (solid red), $\langle a \rangle \langle waw \rangle_c$ term (solid purple), and $\langle aaww \rangle_c$ term (solid yellow). In all cases, we plot $\langle \tilde{D}_{\ell} \rangle = \ell(\ell+1)\langle \tilde{C}_{\ell} \rangle / (2\pi)$. The reMASTERed result displays significantly better agreement with the directly computed power spectrum of the masked map than the MASTER result. }
    \label{fig:ISW_master}
\end{figure}

Figure \ref{fig:ISW_master} shows the the results of our implementation of Eq.~\eqref{eq.full_result_with_rho} on this example, with all correlators computed from a single realization. We use a single realization simply to demonstrate correctness of the analytic results and computational implementation via comparison with the ``directly computed" pseudo-$C_\ell$. In practice, we would perform such an analysis on an ensemble-averaged version as described in \S \ref{sec.computational_implementation}, since our goal is to obtain a theoretical model for $\tilde{C}_\ell$ from the theoretical correlators. Note that the original MASTER result is the $\langle aa \rangle \langle ww \rangle$ term on its own, which displays clear disagreement with the direct calculation of $\tilde{C}_{\ell}$ from the masked map. Our reMASTERed result resolves the discrepancy with the addition of the $\langle aw \rangle \langle aw \rangle$, $\langle w \rangle \langle aaw \rangle_c$, $\langle a \rangle \langle waw \rangle_c$, and $\langle aaww \rangle_c$ terms, which are relevant due to the nonzero correlation of the ISW field map with the threshold mask. In particular, we obtain an improvement in the mean absolute percent error from $\approx 30\%$ with the MASTER result to effectively no error with the reMASTERed result.\footnote{To be precise, the reMASTERed result has a mean absolute percent error that is zero to five decimal places for this case.} The mean absolute percent error is found by calculating the absolute percent error at each $\ell\geq 2$ and taking the mean over the multipoles. 

\begin{figure}[H]
    \centering
    \includegraphics[width=0.60\textwidth]{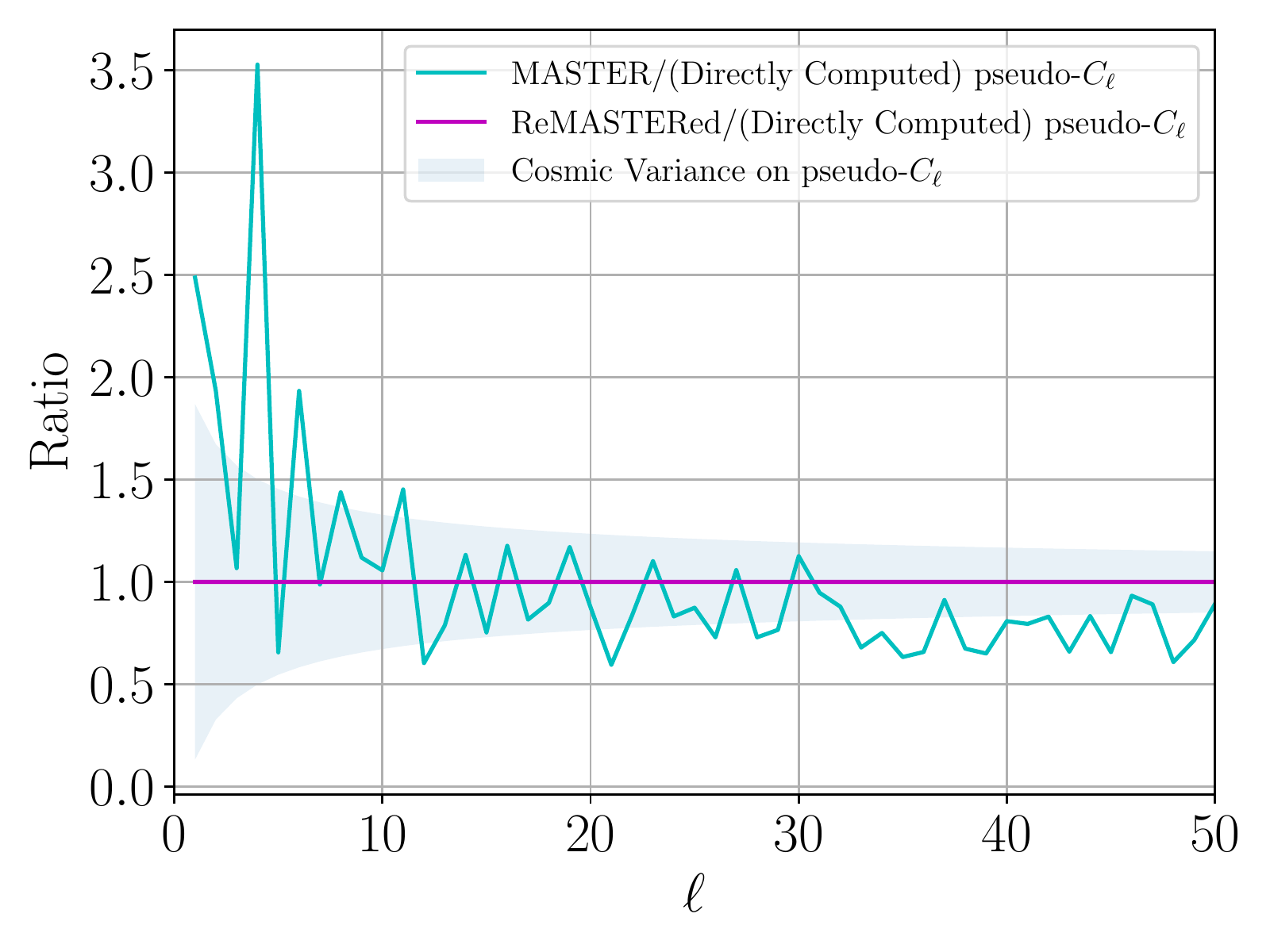}
    \caption{Ratios of original MASTER (solid cyan) and reMASTERed (solid magenta) pseudo-$C_\ell$ to the directly computed pseudo-$C_\ell$ for the ISW field with a threshold mask. The region between the bounds of error due to (the approximate) cosmic variance of the directly computed pseudo-$C_\ell$ is shaded in light blue. }
    \label{fig:ratios_ISW}
\end{figure}

Figure \ref{fig:ratios_ISW} compares ratios of the original MASTER and our reMASTERed results to the directly computed pseudo-$C_\ell$ for this example. For comparison, error bars for the cosmic variance of the pseudo-$C_\ell$ are also shown.\footnote{Here we use an approximation to the pseudo-$C_\ell$ variance, dropping the non-Gaussian contributions. As will be discussed in \S \ref{sec.discussion}, the full covariance is more difficult to compute, and beyond the remit of this paper.} Letting $\tilde{C}_\ell$ be the directly computed pseudo-$C_\ell$, (the approximation to) the cosmic variance is calculated as $\sigma^2_{\tilde{C}_\ell} = \frac{2}{(2\ell+1)f_{\rm sky}} \tilde{C}_\ell^2$, where $f_{\rm sky}$ is the unmasked sky fraction, and the bounds of the cosmic variance on the ratio plot are $(\tilde{C}_\ell \pm \sigma_{\tilde{C}_\ell})/(\tilde{C}_\ell)$. These ratios are useful for assessing the accuracy of the forward-modeled pseudo-$C_\ell$ in the MASTER versus reMASTERed cases. However, as will be discussed in further detail in \S \ref{sec.discussion}, in the MASTER formalism, the mode-coupling matrix is usually inverted and parameter inference is performed using the obtained ``true" $C_\ell$. We explore biases on the true $C_\ell$ obtained via MASTER in Appendix \ref{sec.appendix_true_cl_ratio}.

%%%%%%%%%%%%%%%%%%%%%%%%%%%%%%%%%%%%%%%%%%%%%%%%%%%%%%%%%%%%%%%%%%%
\section{Application: tSZ Field with Masked Infrared Sources}
\label{sec.y_mask_ir}

\begin{figure}[t]
    \centering
    \includegraphics[width=0.9\textwidth]{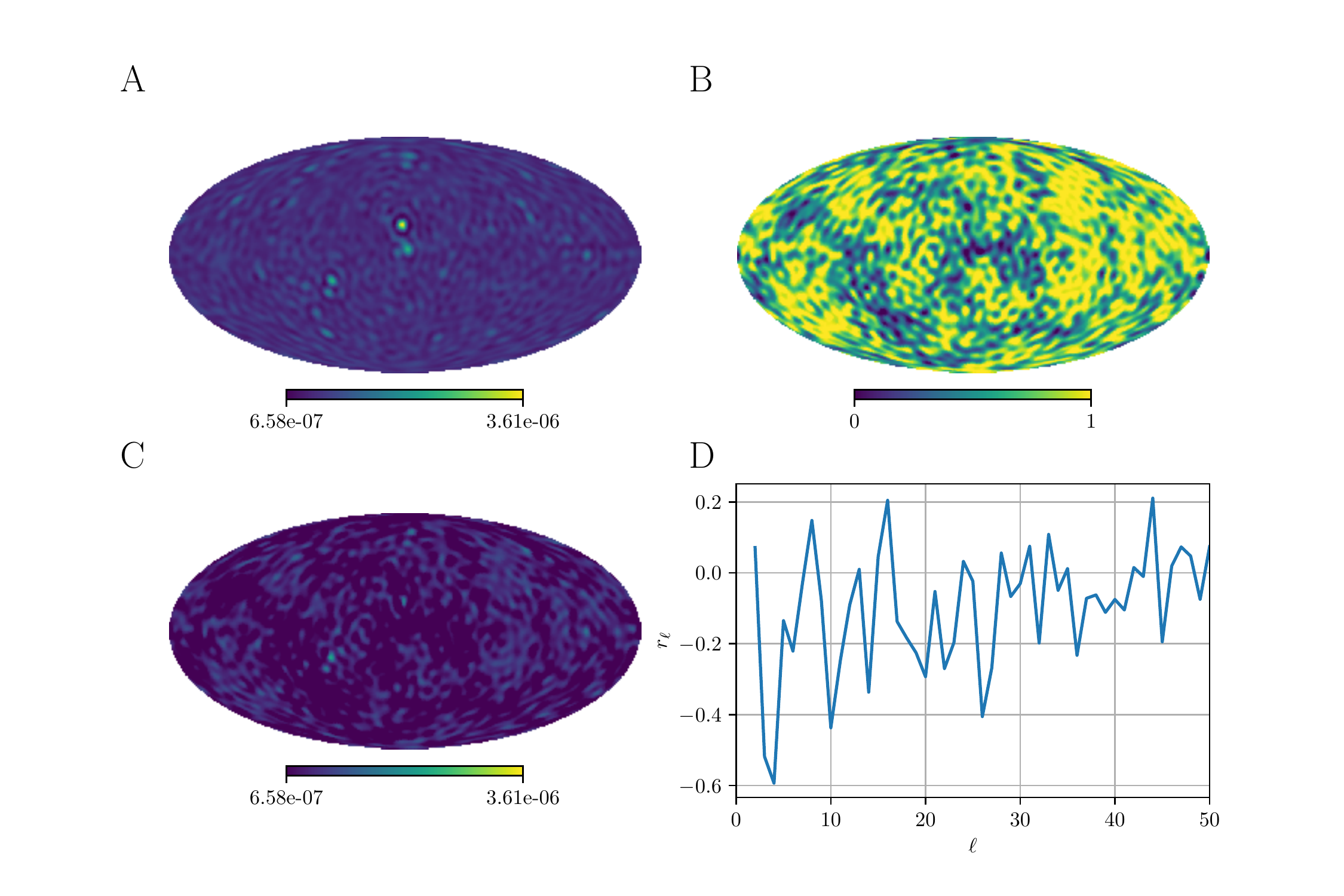}
    \caption{\textbf{A.} WebSky Compton-$y$ map with $N_{\rm side}=128$, in dimensionless Compton-$y$ units. \textbf{B.} Mask designed to mask out infrared sources above $25$ mJy at 153 GHz with 20 arcmin holes, apodized with a cosine apodization of width $2.5^{\circ}$. \textbf{C.} Masked Compton-$y$ map in dimensionless Compton-$y$ units. \textbf{D.} Correlation coefficient of Compton-$y$ map and IR source mask. As in Fig. \ref{fig:ISW_maps}, we find non-trivial correlations at low-$\ell$.}
    \label{fig:tSZ_maskIR_maps}
\end{figure}

The thermal Sunyaev-Zel'dovich (tSZ) effect is the spectral distortion of the CMB blackbody spectrum caused by the inverse-Compton scattering of CMB photons off hot electrons along the line of sight \cite{SZ1969, SZ1970}. To study the tSZ signal, one may often want to mask bright infrared sources to reduce contamination, as in, \textit{e.g.}, Refs.~\cite{Planck:2015tsz, Chiang:2020, wilson2012atacama, Singari:2020}. Because infrared sources are non-negligibly correlated with the tSZ field \cite{addison2012, Planck:2015tSZ_CIB}, the tSZ field with an IR source mask serves as a good example to test our result.

For our implementation, we obtain a Compton-$y$ map and IR source catalog at 153 GHz from WebSky\footnote{We thank Zack Li for providing this catalog, which is publicly available in the ``xzackli" scratch directory on NERSC under the ``cib\_sources" subdirectory. Specifically, we use the files cen\_chunk1\_flux\_153.h5 and cen\_chunk2\_flux\_153.h5 for the flux of central sources and the files sat\_chunk1\_flux\_153.h5 and sat\_chunk2\_flux\_153.h5 for the flux of satellite sources. Moreover, for angle and redshift information about the sources, we use the files cen\_chunk1.h5, cen\_chunk2.h5, sat\_chunk1.h5, and sat\_chunk2.h5, respectively.} \cite{Websky2020}. The WebSky simulations use the CIB model from Ref.~\cite{Viero_2013_hermes}, which used the \emph{Herschel} Multi-tiered Extragalactic Survey (\emph{HerMES}) \cite{Oliver_2012} data from the SPIRE instrument aboard the \emph{Herschel Space Observatory} \cite{Pilbratt_2010} to constrain the standard Shang \textit{et al.} halo model of the CIB emission \cite{Shang_2012}. For the Compton-$y$ map, the authors of Ref.~\cite{Websky2020} projected the pressure profiles determined from the hydrodynamical simulations of Ref.~\cite{battaglia2012cluster} onto the mass-Peak Patch halo catalog \cite{Stein:2018lrh, Bond1996}.

For our use, the $y$-map is downgraded to $N_{\rm side}=128$ for computational speed. A mask is created to mask out IR sources with 20 arcmin holes using a flux cut of $25$ mJy at 153 GHz, resulting in 4377 sources being masked.\footnote{We use holes of this size due to the low $N_{\rm side}$ of the Compton-$y$ map. Keeping a reasonable amount of the sky unmasked prevents us from using a lower flux cut for the IR source mask.} The mask is apodized with a cosine apodization of width $2.5^{\circ}$. Figure \ref{fig:tSZ_maskIR_maps} shows the Compton-$y$ map, IR source mask, and correlation coefficient of the map and mask. Because we only have access to one IR source catalog and corresponding Compton-$y$ field, we are limited to calculating the results for this one realization instead of in the ensemble average for this demonstration. We note that in real data analysis, the correlation coefficient of the Compton-$y$ map and an IR source mask would likely be even higher, and thus the bias from the MASTER approach would be even larger. Reasons for this include the fact that in real data analysis one would likely use a lower flux cut of the IR sources. Moreover, the correlation here is heavily dependent on the CIB model assumed in WebSky.

\begin{figure}[t]
    \centering
    \includegraphics[width=0.80\textwidth]{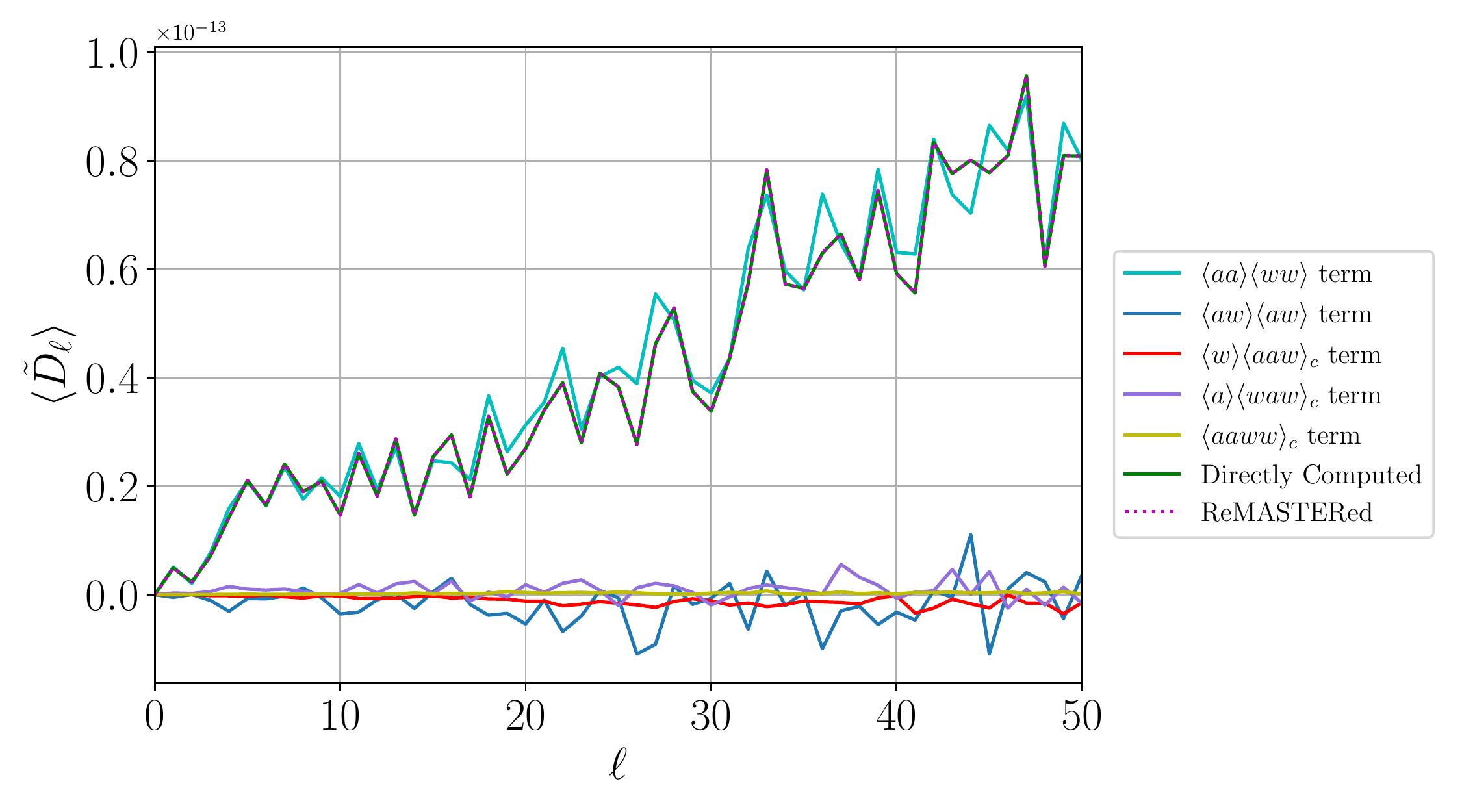}
    \caption{As Fig.~\ref{fig:ISW_master} but for the Compton-$y$ field with an infrared source mask of flux cut $25$ mJy at 153 GHz. $\langle \tilde{D}_{\ell} \rangle$ here is in dimensionless Compton-$y$ units. The reMASTERed result displays significantly better agreement with the directly computed power spectrum than the MASTER result (which corresponds to the $\langle aa \rangle \langle ww \rangle$ term alone).}
    \label{fig:tSZ_maskIR_master}
\end{figure}

\begin{figure}[t]
    \centering
    \includegraphics[width=0.60\textwidth]{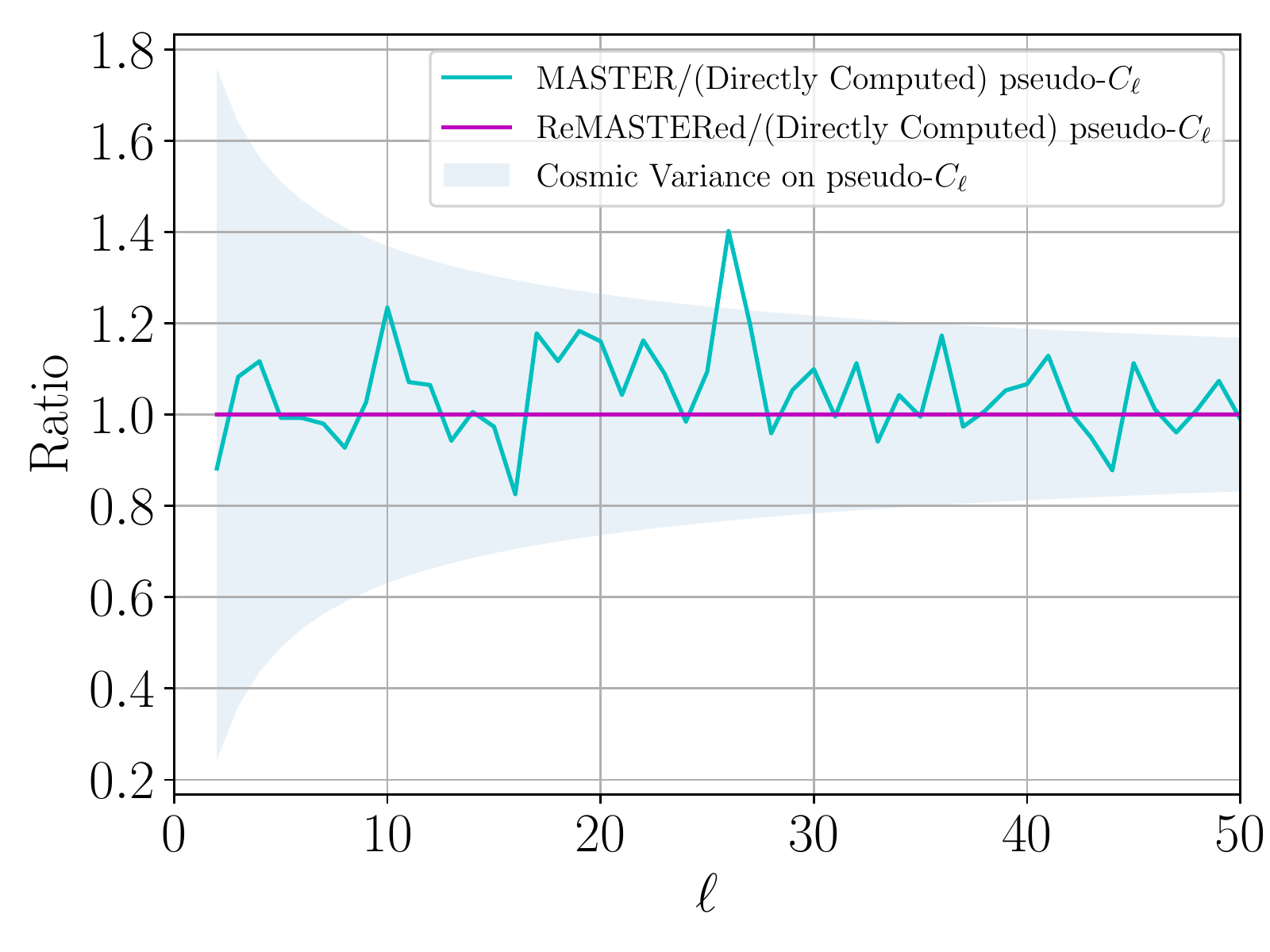}
    \caption{As Fig.~\ref{fig:ratios_ISW} but for the tSZ field with an IR source mask.}
    \label{fig:ratios_tSZ}
\end{figure}

Figure \ref{fig:tSZ_maskIR_master} shows the the results of our implementation of Eq.~\eqref{eq.full_result_with_rho} on this example. Similar to the case for the ISW effect map with a threshold mask, because of the nonzero correlation of the tSZ field with IR sources, the reMASTERed result is a near-perfect match with the directly computed result for the pseudo-$C_\ell$, whereas the original MASTER result displays significant disagreement with it. In particular, we obtain an improvement in the mean absolute percent error from $\approx 10\%$ with the MASTER result to effectively no error with the reMASTERed result. Figure \ref{fig:ratios_tSZ} compares ratios of the original MASTER and our reMASTERed results to the directly computed pseudo-$C_\ell$ for this example, along with the bounds of (the approximation to) the cosmic variance of the pseudo-$C_\ell$.  

As for the ISW field case, we also explore biases on the true $C_\ell$ obtained via inverting the MASTER result in Appendix \ref{sec.appendix_true_cl_ratio}. We note that the tSZ effect is dominant at scales much smaller than those we have considered here. Thus, in Fig.~\ref{fig:high_res_tsz_mask_IR} we show a high-resolution Compton-$y$ map with IR sources down to 5 mJy masked, along with the correlation coefficient of the map and mask out to $\ell = 3000$. In Fig.~\ref{fig:ratio_true_cl_tsz}, we show biases from the MASTER approach out to $\ell=1000$ for this case; we note that these biases are non-negligible even at small scales.

%%%%%%%%%%%%%%%%%%%%%%%%%%%%%%%%%%%%%%%%%%%%%%%%%%%%%%%%%%%%%%%%%%%
\section{Discussion}
\label{sec.discussion}

In this work we have developed an analytic model for the pseudo-$C_\ell$ of maps with correlated masks. In doing so, we found correction terms to the widely used MASTER result for pseudo-$C_\ell$ estimation. Moreover, we have developed a software tool for calculating the relative contributions from each term, forward-modeling the pseudo-$C_\ell$ from these terms, and comparing the results to directly computed pseudo-$C_\ell$ as a validation test. Using this computational implementation, we showed significant improvement of our reMASTERed result over the traditional MASTER result for example cases of physical interest.

An interesting consequence of our result is that, even for Gaussian random fields, the power spectrum of the masked field in general contains higher-point functions of the map and mask. For most fields and masks of interest, it is difficult to predict \textit{a priori} which of the new terms will be most important, as can be seen by comparing the relative contributions of terms in Figs.~\ref{fig:ISW_master}, \ref{fig:tSZ_maskIR_master}, and \ref{fig:tSZ_w_eq_aplusA} (in Appendix~\ref{sec.appendix_consistency_checks}).

In the traditional MASTER formalism, Eq.~\eqref{eq.MASTER_result} can be rewritten via matrix inversion to obtain $\langle C_{\ell}^{aa} \rangle$ in terms of $\langle \tilde{C}_{\ell} \rangle$. A consequence of our new result with additional terms is that this inversion is no longer easily done, making the inference problem significantly harder. Our result instead requires forward-modeling of theoretical predictions into the observable space (\textit{i.e.}, the pseudo-$C_\ell$).  With the usual MASTER result, one can take a single observation of $\tilde{C}_{\ell}$, apply an inversion matrix to obtain $C_\ell^{aa}$, and then run MCMC methods in the theory space, \textit{i.e.}, MCMC in $(C_\ell^{aa}-C_\ell^{aa, \, \rm theory})$. With the reMASTERed result, one would instead use theoretical correlators of the unmasked map and mask to forward-model $\tilde{C}_\ell^{\rm theory}$ and would then need to run MCMC in $(\tilde{C}_{\ell}^{\rm measured}-\tilde{C}_{\ell}^{\text{theory}})$. A possible alternative is to use simulation-based inference, which requires one to compute several realizations of the field and mask at each step of the MCMC to obtain pseudo-$C_\ell$ to compare with that from data. Depending on the specific physical scenario, generating such maps and masks can be very time-consuming or even intractable; for example, one can imagine having to generate several realizations of a non-Gaussian field and point source masks corresponding to that field. In either approach (forward-modeling higher-point functions or full simulation-based inference), the computational expense is significantly greater than that required for the much simpler MASTER-based analysis.  This is an unfortunate byproduct of the use of masks that are correlated with the underlying field of interest. Nevertheless, an alternate approach would be to use our full result only in the first few steps of an MCMC chain, after which the difference between our full reMASTERed result and the MASTER result can be approximated by some linear bias term. This can likely be done since, as seen in the practical examples given in this work, the usual MASTER term is dominant. We leave studies of this linear bias approximation to future work.

Similar to the original MASTER result, we have shown that our results hold in the \textit{ensemble average}. In a practical application, one would have access to theoretical models for correlators of the map and mask but only a single pseudo-$C_\ell$ observation. One could then propagate the theoretical models according to Eq.~\eqref{eq.full_result_with_trispectrum} to obtain an ensemble-averaged pseudo-$C_\ell$ estimate. However, comparison to the single pseudo-$C_\ell$ observation is still useful, especially at high $\ell$ since the cosmic variance scales as $1/(2\ell+1)$. 

The examples we have discussed are scientifically plausible scenarios where a mask might exhibit significant correlations with the signal map. However, the usefulness of our result is not limited to such cases. The result could be important for high-precision measurements where there are even small correlations of the map and mask. The reMASTERed result picks up fluctuations due to random correlations of the map and mask unlike the MASTER result. One would typically bin the data for using the MASTER result, but this may cause one to miss fluctuations that are actually due to the signal.

A limitation of our result is that, similar to the case with the original MASTER result, the presence of a mask leads to complications in assessing the covariance matrix of a power spectrum estimate since different Fourier modes become correlated, leading to a non-diagonal covariance matrix \cite{efstathiou2004myths, Namaster}. Just as our model for the pseudo-$C_\ell$ is now a full four-point function, the general covariance matrix of the pseudo-$C_\ell$ is now a full eight-point function of the map and mask, which is almost completely intractable to calculate. Fortunately, one can instead use simulations to compute the covariance matrix for just one fixed setting of the model parameters, as is often done in current data analyses.

Another limitation is that our result only holds \textit{exactly} for fields and masks such that the field and the portion of the mask that is correlated with the field are statistically isotropic in the ensemble average. However, one may imagine a case in which there is some mask that is anisotropically correlated with the field, \textit{e.g.}, via some multiplicative operation. In such a case, neither the original MASTER result nor our reMASTERed result would be completely accurate in describing the pseudo-$C_\ell$. While it is possible to accurately analytically model what such a result would look like, it is not very useful due to computational intractability---the Wigner $3j$ symbols would contain nonzero $m$ values, making the sums scale as  $\mathcal{O}(\ell_{\rm max}^9)$ instead of our current $\mathcal{O}(\ell_{\rm max}^4)$ for the $\langle aaww \rangle_c$ term and $\mathcal{O}(\ell_{\rm max}^2)$ for the other terms. Nevertheless, even in such cases, our result is likely to be a very good approximation.

%%%%%%%%%%%%%%%%%%%%%%%%%%%%%%%%%%%%%%%%%%%%%%%%%%%%%%%%%%%%%%%%%%%
\section{Acknowledgements}
\label{sec.acknowledgements}
We thank David Alonso, William Coulton, Giulio Fabbian, Eiichiro Komatsu, David Spergel, and the anonymous referee for helpful comments. This material is based upon work supported by the National Science Foundation Graduate Research Fellowship Program under Grant No. DGE 2036197 (KMS). OHEP is a Junior Fellow of the Simons Society of Fellows and acknowledges support and free-flowing Kombucha from the Simons Foundation. JCH acknowledges support from NSF grant AST-2108536, NASA grant 21-ATP21-0129, DOE grant DE-SC00233966, the Sloan Foundation, and the Simons Foundation. This research used resources of the National Energy Research Scientific Computing Center (NERSC), a U.S. Department of Energy Office of Science User Facility located at Lawrence Berkeley National Laboratory, operated under Contract No. DE-AC02-05CH11231. This research also used computing resources from Columbia University's Shared Research Computing Facility project, which is supported by NIH Research Facility Improvement Grant 1G20RR030893-01, and associated funds from the New York State Empire State Development, Division of Science Technology and Innovation (NYSTAR) Contract C090171, both awarded April 15, 2010. 

%%%%%%%%%%%%%%%%%%%%%%%%%%%%%%%%%%%%%%%%%%%%%%%%%%%%%%%%%%%%%%%%%%%
%%%%%%%%%%%%%%%%%%%%%%%%%%%%%%%%%%%%%%%%%%%%%%%%%%%%%%%%%%%%%%%%%%%
%%%%%%%%%%%%%%%%%%%%%%%%%%%%%%%%%%%%%%%%%%%%%%%%%%%%%%%%%%%%%%%%%%%
%%%%%%%%%%%%%%%%%%%%%%%%%%%%%%%%%%%%%%%%%%%%%%%%%%%%%%%%%%%%%%%%%%%
%%%%%%%%%%%%%%%%%%%%%%%%%%%%%%%%%%%%%%%%%%%%%%%%%%%%%%%%%%%%%%%%%%%
%%%%%%%%%%%%%%%%%%%%%%%%%%%%%%%%%%%%%%%%%%%%%%%%%%%%%%%%%%%%%%%%%%%

\begin{appendices}

\section{Full Derivation}
\label{sec.appendix_full_derivation_iso}

Here we derive the result for the expansion of the masked map power spectrum in terms of $n$-point functions, the results of which are given in \S \ref{sec.analytic_derivation}. The ensemble-averaged pseudo-$C_\ell$, $\langle \tilde{C}_{\ell_1} \rangle$, is given by
\begin{align}
    \langle \tilde{C}_{\ell_1} \rangle &\equiv \frac{1}{2\ell_1+1}\sum_{m_1=-\ell_1}^{\ell_1} \langle \tilde{a}_{\ell_1m_1}, \tilde{a}^{*}_{\ell_1m_1} \rangle \nonumber
    \\&=\frac{1}{2\ell_1+1}\sum_{m_1=-\ell_1}^{\ell_1}\sum_{\ell_2 m_2}\sum_{\ell_3 m_3} \langle K_{\ell_1 m_1 \ell_2 m_2}[W] K^{*}_{\ell_1 m_1 \ell_3 m_3}[W]  a_{\ell_2 m_2}a_{\ell_3 m_3}^* \rangle \nonumber
    \\&= \frac{1}{2\ell_1+1}\sum_{m_1=-\ell_1}^{\ell_1}\sum_{\ell_2 m_2}\sum_{\ell_3 m_3} \sum_{\ell_4 m_4} \sum_{\ell_5 m_5} \langle a_{\ell_2 m_2}a^{*}_{\ell_3 m_3} w_{\ell_4 m_4} w^{*}_{\ell_5 m_5} \rangle \mathcal{G}^{\ell_1 \ell_2 \ell_4}_{-m_1 m_2 m_4} \mathcal{G}^{\ell_1 \ell_3 \ell_5}_{-m_1 m_3 m_5} \nonumber
    \\&= \frac{1}{2\ell_1+1}\sum_{m_1=-\ell_1}^{\ell_1}\sum_{\ell_2 m_2}\sum_{\ell_3 m_3} \sum_{\ell_4 m_4} \sum_{\ell_5 m_5} \langle a_{\ell_2 m_2}a^{*}_{\ell_3 m_3} w_{\ell_4 m_4} w^{*}_{\ell_5 m_5} \rangle \mathcal{G}^{\ell_1 \ell_2 \ell_4}_{-m_1 m_2 m_4} \mathcal{G}^{\ell_1 \ell_3 \ell_5}_{m_1 -m_3 -m_5}  \nonumber
    \\&= \frac{1}{2\ell_1+1}\sum_{m_1=-\ell_1}^{\ell_1}\sum_{\ell_2 m_2}\sum_{\ell_3 m_3} \sum_{\ell_4 m_4} \sum_{\ell_5 m_5} (-1)^{m_1} \langle a_{\ell_2 m_2}a_{\ell_3 m_3} w_{\ell_4 m_4} w_{\ell_5 m_5} \rangle \mathcal{G}^{\ell_1 \ell_2 \ell_4}_{-m_1 m_2 m_4} \mathcal{G}^{\ell_1 \ell_3 \ell_5}_{m_1 m_3 m_5} \nonumber
    \\&= \frac{1}{2\ell_1+1} \sum_{m_1=-\ell_1}^{\ell_1}\sum_{\ell_2 m_2}\sum_{\ell_3 m_3} \sum_{\ell_4 m_4} \sum_{\ell_5 m_5} (-1)^{m_1} \mathcal{G}^{\ell_1 \ell_2 \ell_4}_{-m_1 m_2 m_4} \mathcal{G}^{\ell_1 \ell_3 \ell_5}_{m_1 m_3 m_5} \nonumber
    [ \langle a_{\ell_2 m_2}a_{\ell_3m_3}\rangle  \langle w_{\ell_4 m_4}w_{\ell_5m_5}\rangle \notag\\ &\qquad + \langle a_{\ell_2 m_2}w_{\ell_4m_4} \rangle \langle a_{\ell_3 m_3}w_{\ell_5m_5} \rangle + \langle a_{\ell_2 m_2}w_{\ell_5m_5} \rangle \langle a_{\ell_3 m_3}w_{\ell_4m_4} \rangle + \langle w_{\ell_4 m_4} \rangle \langle a_{\ell_2 m_2} a_{\ell_3 m_3} w_{\ell_5 m_5} \rangle_c \nonumber \notag\\ &\qquad + \langle w_{\ell_5 m_5} \rangle \langle a_{\ell_2 m_2} a_{\ell_3 m_3} w_{\ell_4 m_4} \rangle_c + \langle a_{\ell_2 m_2} \rangle \langle a_{\ell_3 m_3} w_{\ell_4 m_4} w_{\ell_5 m_5} \rangle_c + \langle a_{\ell_3 m_3} \rangle \langle a_{\ell_2 m_2} w_{\ell_4 m_4} w_{\ell_5 m_5}\rangle_c \nonumber \notag\\ &\qquad  + \langle a_{\ell_2 m_2} a_{\ell_3 m_3} w_{\ell_4 m_4} w_{\ell_5 m_5} \rangle_c],
    \label{eq.appendix_full_expansion}
\end{align}
where in the last line we have written out the Wick contractions of the four-point function. Then, 
\begin{align}
    \langle \tilde{C}_{\ell_1} \rangle &= \frac{1}{2\ell_1+1} \sum_{m_1=-\ell_1}^{\ell_1}\sum_{\ell_2 m_2}\sum_{\ell_3 m_3} \sum_{\ell_4 m_4} \sum_{\ell_5 m_5} (-1)^{m_1} \mathcal{G}^{\ell_1 \ell_2 \ell_4}_{-m_1 m_2 m_4} \mathcal{G}^{\ell_1 \ell_3 \ell_5}_{m_1 m_3 m_5} \bigg[ 
    \nonumber \notag\\ &\qquad \langle C_{\ell_2}^{aa}\rangle  \langle C_{\ell_4}^{ww}\rangle (-1)^{m_2+m_4} \delta^{\rm K}_{\ell_2, \ell_3}\delta^{\rm K}_{m_2,-m_3}\delta^{\rm K}_{\ell_4, \ell_5}\delta^{\rm K}_{m_4,-m_5} + \langle C_{\ell_2}^{aw} \rangle \langle C_{\ell_3}^{aw} \rangle (-1)^{m_2+m_3} \delta^{\rm K}_{\ell_2, \ell_4}\delta^{\rm K}_{m_2,-m_4}\delta^{\rm K}_{\ell_3, \ell_5}\delta^{\rm K}_{m_3,-m_5} 
     \nonumber \notag\\ &\qquad + \langle C_{\ell_2}^{aw} \rangle \langle C_{\ell_3}^{aw} \rangle (-1)^{m_2+m_3} \delta^{\rm K}_{\ell_2, \ell_5}\delta^{\rm K}_{m_2,-m_5}\delta^{\rm K}_{\ell_3, \ell_4}\delta^{\rm K}_{m_3,-m_4} 
    \nonumber \notag\\ &\qquad + \langle w_{\ell_4 m_4} \rangle \mathcal{G}^{\ell_2 \ell_3 \ell_5}_{m_2 m_3 m_5} b_{\ell_2 \ell_3 \ell_5}^{aaw} +  \langle w_{\ell_5 m_5} \rangle \mathcal{G}^{\ell_2 \ell_3 \ell_4}_{m_2 m_3 m_4} b_{\ell_2 \ell_3 \ell_4}^{aaw} + \langle a_{\ell_2 m_2} \rangle G^{\ell_3 \ell_4 \ell_5}_{m_3 m_4 m_5} b^{aww}_{\ell_3 \ell_4 \ell_5} + \langle a_{\ell_3 m_3} \rangle G^{\ell_2 \ell_4 \ell_5}_{m_2 m_4 m_5} b^{aww}_{\ell_2 \ell_4 \ell_5}  \nonumber \notag\\ &\qquad + \sum_{L=0}^{\infty} \sum_{M=-L}^{L} (-1)^M \mathcal{G}^{\ell_2 \ell_3 L}_{m_2 m_3 -M} \mathcal{G}^{\ell_4 \ell_5 L}_{m_4 m_5 M} t[aaww]^{\ell_2 \ell_3}_{\ell_4 \ell_5}(L) + \text{ 23 perms.} \bigg],
\end{align}
where we have assumed isotropy of the mask. Applying the Kronecker deltas and separating the terms:

\begin{align}
    \label{eq.eight_terms}
    \langle \tilde{C}_{\ell_1} \rangle &= \frac{1}{2\ell_1+1} \sum_{m_1=-\ell_1}^{\ell_1}\sum_{\ell_2 m_2}  \sum_{\ell_4 m_4} (-1)^{m_1+m_2+m_4} \mathcal{G}^{\ell_1 \ell_2 \ell_4}_{-m_1 m_2 m_4} \mathcal{G}^{\ell_1 \ell_2 \ell_4}_{m_1 -m_2 -m_4} \langle C_{\ell_2}^{aa}\rangle  \langle C_{\ell_4}^{ww}\rangle  
    \\\nonumber \notag &\;\; + \frac{1}{2\ell_1+1} \sum_{m_1=-\ell_1}^{\ell_1}\sum_{\ell_2 m_2} \sum_{\ell_3 m_3} (-1)^{m_1+m_2+m_3} \mathcal{G}^{\ell_1 \ell_2 \ell_2}_{-m_1 m_2 -m_2} \mathcal{G}^{\ell_1 \ell_3 \ell_3}_{m_1 m_3 -m_3}   \langle C_{\ell_2}^{aw}\rangle  \langle C_{\ell_3}^{aw}\rangle     
    \nonumber \notag\\ &\;\; + \frac{1}{2\ell_1 + 1} \sum_{m_1=-\ell_1}^{\ell_1}\sum_{\ell_2 m_2} \sum_{\ell_3 m_3} (-1)^{m_1+m_2+m_3} \mathcal{G}^{\ell_1 \ell_2 \ell_3}_{-m_1 m_2 -m_3} \mathcal{G}^{\ell_1 \ell_3 \ell_2}_{m_1 m_3 -m_2} \langle C_{\ell_2}^{aw}\rangle  \langle C_{\ell_3}^{aw}\rangle 
    \nonumber \notag\\ &\;\; + \frac{1}{2\ell_1+1} \sum_{m_1=-\ell_1}^{\ell_1}\sum_{\ell_2 m_2}\sum_{\ell_3 m_3} \sum_{\ell_4 m_4} \sum_{\ell_5 m_5} (-1)^{m_1} \mathcal{G}^{\ell_1 \ell_2 \ell_4}_{-m_1 m_2 m_4} \mathcal{G}^{\ell_1 \ell_3 \ell_5}_{m_1 m_3 m_5} \mathcal{G}^{\ell_2 \ell_3 \ell_5}_{m_2 m_3 m_5} \langle w_{\ell_4 m_4} \rangle  \langle b_{\ell_2 \ell_3 \ell_5}^{aaw} \rangle 
    \nonumber \notag\\ &\;\; + \frac{1}{2\ell_1+1} \sum_{m_1=-\ell_1}^{\ell_1}\sum_{\ell_2 m_2}\sum_{\ell_3 m_3} \sum_{\ell_4 m_4} \sum_{\ell_5 m_5} (-1)^{m_1} \mathcal{G}^{\ell_1 \ell_2 \ell_4}_{-m_1 m_2 m_4} \mathcal{G}^{\ell_1 \ell_3 \ell_5}_{m_1 m_3 m_5} \mathcal{G}^{\ell_2 \ell_3 \ell_4}_{m_2 m_3 m_4} \langle w_{\ell_5 m_5} \rangle  \langle b_{\ell_2 \ell_3 \ell_4}^{aaw} \rangle 
    \nonumber \notag\\ &\;\; + \frac{1}{2\ell_1+1} \sum_{m_1=-\ell_1}^{\ell_1}\sum_{\ell_2 m_2}\sum_{\ell_3 m_3} \sum_{\ell_4 m_4} \sum_{\ell_5 m_5} (-1)^{m_1} \mathcal{G}^{\ell_1 \ell_2 \ell_4}_{-m_1 m_2 m_4} \mathcal{G}^{\ell_1 \ell_3 \ell_5}_{m_1 m_3 m_5} \mathcal{G}^{\ell_3 \ell_4 \ell_5}_{m_3 m_4 m_5} \langle a_{\ell_2 m_2} \rangle  \langle b_{\ell_3 \ell_4 \ell_5}^{aww} \rangle 
    \nonumber \notag\\ &\;\; + \frac{1}{2\ell_1+1} \sum_{m_1=-\ell_1}^{\ell_1}\sum_{\ell_2 m_2}\sum_{\ell_3 m_3} \sum_{\ell_4 m_4} \sum_{\ell_5 m_5} (-1)^{m_1} \mathcal{G}^{\ell_1 \ell_2 \ell_4}_{-m_1 m_2 m_4} \mathcal{G}^{\ell_1 \ell_3 \ell_5}_{m_1 m_3 m_5} \mathcal{G}^{\ell_2 \ell_4 \ell_5}_{m_2 m_4 m_5} \langle a_{\ell_3 m_3} \rangle  \langle b_{\ell_2 \ell_4 \ell_5}^{aww} \rangle 
    \nonumber \notag\\ &\;\; + \frac{1}{2\ell_1+1} \sum_{m_1=-\ell_1}^{\ell_1}\sum_{\ell_2 m_2}\sum_{\ell_3 m_3} \sum_{\ell_4 m_4} \sum_{\ell_5 m_5} \sum_{LM} (-1)^{m_1+M} \mathcal{G}^{\ell_1 \ell_2 \ell_4}_{-m_1 m_2 m_4} \mathcal{G}^{\ell_1 \ell_3 \ell_5}_{m_1 m_3 m_5} \mathcal{G}^{\ell_2 \ell_3 L}_{m_2 m_3 -M} \mathcal{G}^{\ell_4 \ell_5 L}_{m_4 m_5 M} t[aaww]^{\ell_2 \ell_3}_{\ell_4 \ell_5}(L) 
    \nonumber \notag\\ &\hspace{390 pt} + \text{ 23 perms.}\nonumber,
    \end{align}
where the 23 permutations are taken over the reduced trispectrum and only the final two Gaunt factors. Schematically, we refer to each of the above terms as $\langle aa \rangle \langle ww \rangle$, $\langle aw \rangle \langle aw \rangle$, $\langle w \rangle \langle aaw \rangle_c$, $\langle a \rangle \langle waw \rangle_c$, and $\langle aaww \rangle_c$. 

We can simplify Eq.~\eqref{eq.eight_terms} using the orthogonality relations and properties of the Wigner $3j$ symbols, Wigner $6j$ symbols, and Gaunt integrals.\footnote{\url{https://functions.wolfram.com/HypergeometricFunctions/ThreeJSymbol/}} We will simplify each of the terms in Eq.~\eqref{eq.eight_terms} separately. Consider the $\langle aa \rangle \langle ww \rangle$ term:
\begin{align}
    &\frac{1}{2\ell_1+1} \sum_{m_1=-\ell_1}^{\ell_1}\sum_{\ell_2 m_2} \sum_{\ell_4 m_4} (-1)^{m_1+m_2+m_4} \mathcal{G}^{\ell_1 \ell_2 \ell_4}_{-m_1 m_2 m_4} \mathcal{G}^{\ell_1 \ell_2 \ell_4}_{m_1 -m_2 -m_4} \langle C_{\ell_2}^{aa}\rangle  \langle C_{\ell_4}^{ww}\rangle  \nonumber
    \\&= \frac{1}{4\pi} \sum_{\ell_2,\ell_3} (2\ell_2+1)(2\ell_3+1) \begin{pmatrix} \ell_1&\ell_2&\ell_3 \\ 0&0&0 \end{pmatrix}^2 \langle C_{\ell_2}^{aa}\rangle  \langle C_{\ell_3}^{ww}\rangle \label{eq.term1_lastline} \,.
\end{align}
This is the original MASTER result. Next consider the first $\langle aw \rangle \langle aw \rangle$ term from Eq.~\eqref{eq.eight_terms}:
\begin{align}
    &\frac{1}{2\ell_1+1} \sum_{m_1=-\ell_1}^{\ell_1}\sum_{\ell_2 m_2} \sum_{\ell_3 m_3} (-1)^{m_1+m_2+m_3} \mathcal{G}^{\ell_1 \ell_2 \ell_2}_{-m_1 m_2 -m_2} \mathcal{G}^{\ell_1 \ell_3 \ell_3}_{m_1 m_3 -m_3}  \langle C_{\ell_2}^{aw}\rangle  \langle C_{\ell_3}^{aw}\rangle  \nonumber
    \\&= \frac{1}{2\ell_1+1} \sum_{\ell_2 m_2} \sum_{\ell_3 m_3} (-1)^{m_2+m_3} \mathcal{G}^{\ell_1 \ell_2 \ell_2}_{0 m_2 -m_2} \mathcal{G}^{\ell_1 \ell_3 \ell_3}_{0 m_3 -m_3}  \langle C_{\ell_2}^{aw}\rangle  \langle C_{\ell_3}^{aw}\rangle \nonumber
    \\&=0 \text{ 
 unless $\ell_1=0$}\label{eq.term2_lastline} \,.
\end{align} 
The second $\langle aw \rangle \langle aw \rangle$ term in Eq.~\eqref{eq.eight_terms} becomes:
\begin{align}
    &\frac{1}{2\ell_1+1} \sum_{m_1=-\ell_1}^{\ell_1}\sum_{\ell_2 m_2} \sum_{\ell_3 m_3}  (-1)^{m_1+m_2+m_3} \mathcal{G}^{\ell_1 \ell_2 \ell_3}_{-m_1 m_2 -m_3}  \mathcal{G}^{\ell_1 \ell_3 \ell_2}_{m_1 m_3 -m_2} \langle C_{\ell_2}^{aw}\rangle  \langle C_{\ell_3}^{aw}\rangle \nonumber
    \\&= \frac{1}{4\pi} \sum_{\ell_2,\ell_3} (2\ell_2+1)(2\ell_3+1) \begin{pmatrix} \ell_1&\ell_2&\ell_3 \\ 0&0&0 \end{pmatrix}^2 \langle C_{\ell_2}^{aw}\rangle  \langle C_{\ell_3}^{aw}\rangle \label{eq.term3_lastline} \,.
\end{align}
The first $\langle w \rangle \langle aaw \rangle_c$ term in Eq.~\eqref{eq.eight_terms} becomes:
\begin{align}
    &\frac{1}{2\ell_1+1} \sum_{m_1=-\ell_1}^{\ell_1}\sum_{\ell_2 m_2}\sum_{\ell_3 m_3} \sum_{\ell_4 m_4} \sum_{\ell_5 m_5}   (-1)^{m_1}\mathcal{G}^{\ell_1 \ell_2 \ell_4}_{-m_1 m_2 m_4} \mathcal{G}^{\ell_1 \ell_3 \ell_5}_{m_1 m_3 m_5} \mathcal{G}^{\ell_2 \ell_3 \ell_5}_{m_2 m_3 m_5} \langle w_{\ell_4 m_4} \rangle  \langle b_{\ell_2 \ell_3 \ell_5}^{aaw} \rangle \nonumber
    \\&=\frac{1}{(4\pi)^{3/2}} \sum_{\ell_2,\ell_3} (2\ell_2+1)(2\ell_3+1) \begin{pmatrix} \ell_1&\ell_2&\ell_3 \\ 0&0&0 \end{pmatrix}^2 \langle w_{00} \rangle  \langle b_{\ell_1 \ell_2 \ell_3}^{aaw} \rangle \label{eq.term4_lastline} \,.
\end{align}
We note that the fact that only $\langle w_{00} \rangle$ contributes is forced by the Gaunt factors. This is relevant to our discussion of anisotropic components of the mask in \S \ref{sec.analytic_derivation}. The next $\langle w \rangle \langle aaw \rangle_c$ term and the $\langle a \rangle \langle waw \rangle_c$ terms in Eq.~\eqref{eq.eight_terms} can be simplified similarly. Finally, consider the $\langle aaww \rangle_c$ term in Eq.~\eqref{eq.eight_terms}. Comparing to Eq.~\eqref{eq.rho_expectation}, we see that this term is just
\begin{equation}
    \frac{1}{2\ell_1+1} \sum_{\ell_2 \ell_3 \ell_4 \ell_5} \langle \hat{\rho}[awaw]^{\ell_2 \ell_4}_{\ell_3 \ell_5}(\ell_1) \rangle \,.
\end{equation}

Alternatively, we may wish to model the final term in terms of the reduced trispectrum, in which case there are three types of permutations we must consider: permutations where $\ell_2$ remains in the same Gaunt integral as $\ell_3$ and $\ell_4$ remains in the same Gaunt integral as $\ell_5$ in Eq.~\eqref{eq.trispectrum_def}; permutations where $\ell_2$ remains in the same Gaunt integral as $\ell_5$ and $\ell_3$ remains in the same Gaunt integral as $\ell_4$ in Eq.~\eqref{eq.trispectrum_def}; and permutations where $\ell_2$ remains in the same Gaunt integral as $\ell_4$ and $\ell_3$ remains in the same Gaunt integral as $\ell_5$ in Eq.~\eqref{eq.trispectrum_def}. Under the permutation $\mathcal{G}^{\ell_2 \ell_3 L}_{m_2 m_3 -M} \mathcal{G}^{\ell_4 \ell_5 L}_{m_4 m_5 M} t[aaww]^{\ell_2 \ell_3}_{\ell_4 \ell_5}(L)$ we find 8 permutations of the form:
\begin{align}
     &\frac{1}{2\ell_1+1} \sum_{m_1=-\ell_1}^{\ell_1}\sum_{\ell_2 m_2}\sum_{\ell_3 m_3} \sum_{\ell_4 m_4} \sum_{\ell_5 m_5} \sum_{LM} (-1)^{m_1+M} \mathcal{G}^{\ell_1 \ell_2 \ell_4}_{-m_1 m_2 m_4} \mathcal{G}^{\ell_1 \ell_3 \ell_5}_{m_1 m_3 m_5} \mathcal{G}^{\ell_2 \ell_3 L}_{m_2 m_3 -M} \mathcal{G}^{\ell_4 \ell_5 L}_{m_4 m_5 M} t[aaww]^{\ell_2 \ell_3}_{\ell_4 \ell_5}(L) \nonumber
      \\&= \frac{1}{(4\pi)^2} \sum_{\ell_2 \ell_3 \ell_4 \ell_5 L} (-1)^{\ell_1+L} (2\ell_2+1)(2\ell_3+1)(2\ell_4+1)(2\ell_5+1)(2L+1) \begin{Bmatrix} \ell_4&\ell_2&\ell_1 \\ \ell_3&\ell_5&L \end{Bmatrix} \nonumber \notag\\ &\qquad \times  \begin{pmatrix} \ell_1&\ell_2&\ell_4 \\ 0&0&0 \end{pmatrix} \begin{pmatrix} \ell_1&\ell_3&\ell_5 \\ 0&0&0 \end{pmatrix} \begin{pmatrix} \ell_2&\ell_3&L \\ 0&0&0 \end{pmatrix} \begin{pmatrix} \ell_4&\ell_5&L \\ 0&0&0 \end{pmatrix} t[aaww]^{\ell_2 \ell_3}_{\ell_4 \ell_5}(L) \label{eq.aaww_term_perm1}
\end{align}
Under the permutation $\mathcal{G}^{\ell_2 \ell_5 L}_{m_2 m_5 -M} \mathcal{G}^{\ell_3 \ell_4 L}_{m_3 m_4 M} t[awaw]^{\ell_2 \ell_5}_{\ell_3 \ell_4}(L)$, there are another 8 analogous terms:
\begin{align}
     &\frac{1}{2\ell_1+1} \sum_{m_1=-\ell_1}^{\ell_1}\sum_{\ell_2 m_2}\sum_{\ell_3 m_3} \sum_{\ell_4 m_4} \sum_{\ell_5 m_5} \sum_{LM} (-1)^{m_1+M} \mathcal{G}^{\ell_1 \ell_2 \ell_4}_{-m_1 m_2 m_4} \mathcal{G}^{\ell_1 \ell_3 \ell_5}_{m_1 m_3 m_5} \mathcal{G}^{\ell_2 \ell_5 L}_{m_2 m_5 -M} \mathcal{G}^{\ell_3 \ell_4 L}_{m_3 m_4 M} t[awaw]^{\ell_2 \ell_4}_{\ell_3 \ell_5}(L) \nonumber
      \\&= \frac{1}{(4\pi)^2} \sum_{\ell_2 \ell_3 \ell_4 \ell_5 L} (-1)^{\ell_1+L} (2\ell_2+1)(2\ell_3+1)(2\ell_4+1)(2\ell_5+1)(2L+1) \begin{Bmatrix} \ell_4&\ell_2&\ell_1 \\ \ell_5&\ell_3&L \end{Bmatrix} \nonumber \notag\\ &\qquad \times  \begin{pmatrix} \ell_1&\ell_2&\ell_4 \\ 0&0&0 \end{pmatrix} \begin{pmatrix} \ell_1&\ell_3&\ell_5 \\ 0&0&0 \end{pmatrix} \begin{pmatrix} \ell_3&\ell_4&L \\ 0&0&0 \end{pmatrix}  \begin{pmatrix} \ell_2&\ell_5&L \\ 0&0&0 \end{pmatrix} t[awaw]^{\ell_2 \ell_5}_{\ell_3 \ell_4}(L) \label{eq.aaww_term_perm2}
\end{align} Finally, under the permutation $\mathcal{G}^{\ell_3 \ell_5 L}_{m_3 m_5 -M} \mathcal{G}^{\ell_2 \ell_4 L}_{m_2 m_4 M} t[awaw]^{\ell_3 \ell_5}_{\ell_2 \ell_4}(L)$ we have 8 terms of the form:
\begin{align}
    &\frac{1}{2\ell_1+1} \sum_{m_1=-\ell_1}^{\ell_1}\sum_{\ell_2 m_2}\sum_{\ell_3 m_3} \sum_{\ell_4 m_4} \sum_{\ell_5 m_5} \sum_{LM} (-1)^{m_1+M} \mathcal{G}^{\ell_1 \ell_2 \ell_4}_{-m_1 m_2 m_4} \mathcal{G}^{\ell_1 \ell_3 \ell_5}_{m_1 m_3 m_5} \mathcal{G}^{\ell_3 \ell_5 L}_{m_3 m_5 -M} \mathcal{G}^{\ell_2 \ell_4 L}_{m_2 m_4 M} t[awaw]^{\ell_3 \ell_5}_{\ell_2 \ell_4}(L) \nonumber
    \\&= \frac{1}{(4\pi)^2} \sum_{\ell_2 \ell_3 \ell_4 \ell_5} (2\ell_2+1)(2\ell_3+1)(2\ell_4+1)(2\ell_5+1) \begin{pmatrix} \ell_1&\ell_2&\ell_4 \\ 0&0&0 \end{pmatrix}^2 \begin{pmatrix} \ell_1&\ell_3&\ell_5 \\ 0&0&0 \end{pmatrix}^2 t[awaw]^{\ell_3 \ell_5}_{\ell_2 \ell_4}(\ell_1) \label{eq.aaww_term_perm3}
\end{align} 

Combining the results for each of the terms, we have
\begin{align}
     \langle \tilde{C}_{\ell_1} \rangle 
    &= \frac{1}{4\pi} \sum_{\ell_2,\ell_3} (2\ell_2+1)(2\ell_3+1) \begin{pmatrix} \ell_1&\ell_2&\ell_3 \\ 0&0&0 \end{pmatrix}^2 \left[  \langle C_{\ell_2}^{aa}\rangle
    \langle C_{\ell_3}^{ww}\rangle +  \langle C_{\ell_2}^{aw}\rangle  \langle C_{\ell_3}^{aw}\rangle + \frac{\langle w_{00} \rangle}{\sqrt{\pi}}  \langle b_{\ell_1 \ell_2 \ell_3}^{aaw} \rangle + \frac{\langle a_{00} \rangle}{\sqrt{\pi}}  \langle b_{\ell_1 \ell_2 \ell_3}^{waw} \rangle  \right] \notag \nonumber \\&\qquad + \frac{1}{2\ell_1+1} \sum_{\ell_2 \ell_3 \ell_4 \ell_5} \langle \hat{\rho}[awaw]^{\ell_2 \ell_4}_{\ell_3 \ell_5}(\ell_1) \rangle
\end{align}
in terms of $\hat{\rho}$, or, in terms of the reduced trispectrum, 

\begin{align}
    \langle \tilde{C}_{\ell_1} \rangle 
    &= \frac{1}{4\pi} \sum_{\ell_2,\ell_3} (2\ell_2+1)(2\ell_3+1) \begin{pmatrix} \ell_1&\ell_2&\ell_3 \\ 0&0&0 \end{pmatrix}^2 \left[  \langle C_{\ell_2}^{aa}\rangle
    \langle C_{\ell_3}^{ww}\rangle +  \langle C_{\ell_2}^{aw}\rangle  \langle C_{\ell_3}^{aw}\rangle + \frac{\langle w_{00} \rangle}{\sqrt{\pi}}  \langle b_{\ell_1 \ell_2 \ell_3}^{aaw} \rangle + \frac{\langle a_{00} \rangle}{\sqrt{\pi}}  \langle b_{\ell_1 \ell_2 \ell_3}^{waw} \rangle  \right] \\\notag \nonumber &+ \frac{8}{(4\pi)^2} \sum_{\ell_2 \ell_3 \ell_4 \ell_5} (2\ell_2+1)(2\ell_3+1)(2\ell_4+1)(2\ell_5+1) \begin{pmatrix} \ell_1&\ell_2&\ell_4 \\ 0&0&0 \end{pmatrix} \begin{pmatrix} \ell_1&\ell_3&\ell_5 \\ 0&0&0 \end{pmatrix} \notag \nonumber \\&\times \Bigg[ \sum_L (-1)^{\ell_1+L}  \begin{Bmatrix} \ell_4&\ell_2&\ell_1 \\ \ell_3&\ell_5&L \end{Bmatrix} \begin{pmatrix} \ell_2&\ell_3&L \\ 0&0&0 \end{pmatrix} \begin{pmatrix} \ell_4&\ell_5&L \\ 0&0&0 \end{pmatrix} t[aaww]^{\ell_2 \ell_3}_{\ell_4 \ell_5}(L)  \notag \nonumber \\&\;\;\; + \sum_L (-1)^{\ell_1+L}  \begin{Bmatrix} \ell_4&\ell_2&\ell_1 \\ \ell_5&\ell_3&L \end{Bmatrix} \begin{pmatrix} \ell_2&\ell_5&L \\ 0&0&0 \end{pmatrix} \begin{pmatrix} \ell_3&\ell_4&L \\ 0&0&0 \end{pmatrix} t[awaw]^{\ell_2 \ell_5}_{\ell_3 \ell_4}(L) \notag \nonumber \\&\;\;\;+ \begin{pmatrix} \ell_1&\ell_2&\ell_4 \\ 0&0&0 \end{pmatrix} \begin{pmatrix} \ell_1&\ell_3&\ell_5 \\ 0&0&0 \end{pmatrix} t[awaw]^{\ell_3 \ell_5}_{\ell_2 \ell_4}(\ell_1) \Bigg]\nonumber.
\end{align}

%%%%%%%%%%%%%%%%%%%%%%%%%%%%%%%%%%%%%%%%%%%%%%%%%%%%%%%%%%%%%%
%%%%%%%%%%%%%%%%%%%%%%%%%%%%%%%%%%%%%%%%%%%%%%%%%%%%%%%%%%%%%%
\section{Consistency Checks}
\label{sec.appendix_consistency_checks}
Here, we discuss various consistency checks performed to assess the validity of our results as well as other test cases for demonstration purposes.

\begin{figure}[htbp]
    \centering
    \includegraphics[width=0.80\textwidth]{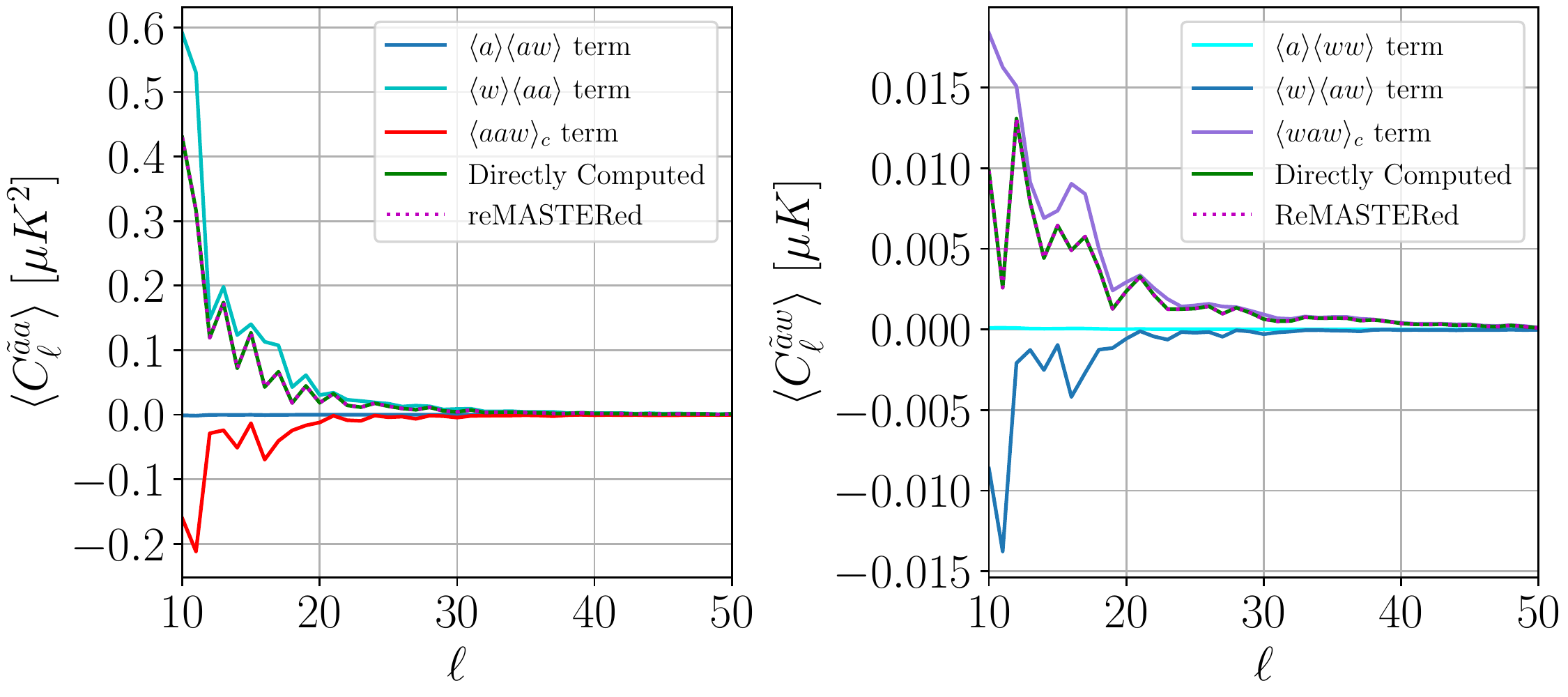}
    \caption{Evaluation of Eq.~\eqref{eq.atildea} for $\langle C_{\ell}^{\tilde{a}a} \rangle$ and Eq.~\eqref{eq.wtildea} for $\langle C_{\ell}^{\tilde{a}w} \rangle$ for the ISW field with threshold mask described in \S \ref{sec.isw_threshold}. The directly computed power spectrum (solid green) is compared to our reMASTERed result (dotted magenta). The agreement implies the validity of our results and implementation, up to the trispectrum term.}
    \label{fig:consistency_ISW}
\end{figure}

We can check terms up to the three-point functions with tests of $C_{\ell}^{\tilde{a} a}$ and $C_{\ell}^{\tilde{a} w}$. In particular,
\begin{equation}
    \label{eq.atildea}
    \langle C_{\ell_1}^{\tilde{a} a} \rangle = \frac{1}{4\pi} \sum_{\ell_2,\ell_3} (2\ell_2+1)(2\ell_3+1) \begin{pmatrix} \ell_1&\ell_2&\ell_3 \\ 0&0&0 \end{pmatrix}^2 \langle b^{aaw}_{\ell_1 \ell_2 \ell_3} \rangle  + \frac{\langle w_{00} \rangle}{\sqrt{4\pi}} \langle C_{\ell_1}^{aa} \rangle + \frac{\langle a_{00} \rangle}{\sqrt{4\pi}} \langle C_{\ell_1}^{aw} \rangle
\end{equation}
and
\begin{equation}
    \label{eq.wtildea}
    \langle C_{\ell_1}^{\tilde{a} w} \rangle = \frac{1}{4\pi} \sum_{\ell_2,\ell_3} (2\ell_2+1)(2\ell_3+1) \begin{pmatrix} \ell_1&\ell_2&\ell_3 \\ 0&0&0 \end{pmatrix}^2 \langle b^{waw}_{\ell_1 \ell_2 \ell_3} \rangle + \frac{\langle w_{00} \rangle}{\sqrt{4\pi}} \langle C_{\ell_1}^{aw} \rangle + \frac{\langle a_{00} \rangle}{\sqrt{4\pi}} \langle C_{\ell_1}^{ww} \rangle \,.
\end{equation}
Using the same schematic argument as in \S \ref{sec.analytic_derivation} to determine which quantities have their ensemble averages removed, 
\begin{equation}
    \langle C_{\ell_1}^{\tilde{a}a} \rangle \sim \langle \tilde{a} a \rangle = \underbrace{\langle a \rangle \langle a \rangle \langle w \rangle + \langle w \rangle \langle \delta a \delta a \rangle}_{\sim \langle w \rangle C_{\ell}^{aa}} + \underbrace{2\langle a \rangle \langle \delta a \delta w \rangle}_{\sim \langle a \rangle C_{\ell}^{\delta a \delta w }} + \underbrace{\langle \delta a \delta a \delta w \rangle_c}_{\sim \mathrm{bispectrum \; term}}
\end{equation}
and
\begin{equation}
    \langle C_{\ell_1}^{\tilde{a}w} \rangle \sim \langle \tilde{a} w \rangle = \underbrace{\langle a \rangle \langle w \rangle \langle w \rangle + \langle a \rangle \langle \delta w \delta w \rangle}_{\sim \langle a \rangle C_{\ell}^{ww}} + \underbrace{2 \langle w \rangle \langle \delta a \delta w \rangle }_{\sim \langle w \rangle C_{\ell}^{\delta a \delta w}} + \underbrace{\langle \delta w \delta a \delta w \rangle_c}_{\sim \mathrm{bispectrum \; term}} \,.
\end{equation}
Thus, the ensemble averages of the map and mask must be subtracted in the $C_{\ell}^{aw}$ and in the bispectra. Figure \ref{fig:consistency_ISW} demonstrates the validity of these results for the ISW field with threshold mask described in \S \ref{sec.isw_threshold}. Specifically, it demonstrates the validity of our bispectrum estimator by allowing us to neglect any contributions that would come from the trispectrum.

\begin{figure}[H]
    \centering
    \includegraphics[width=0.80\textwidth]{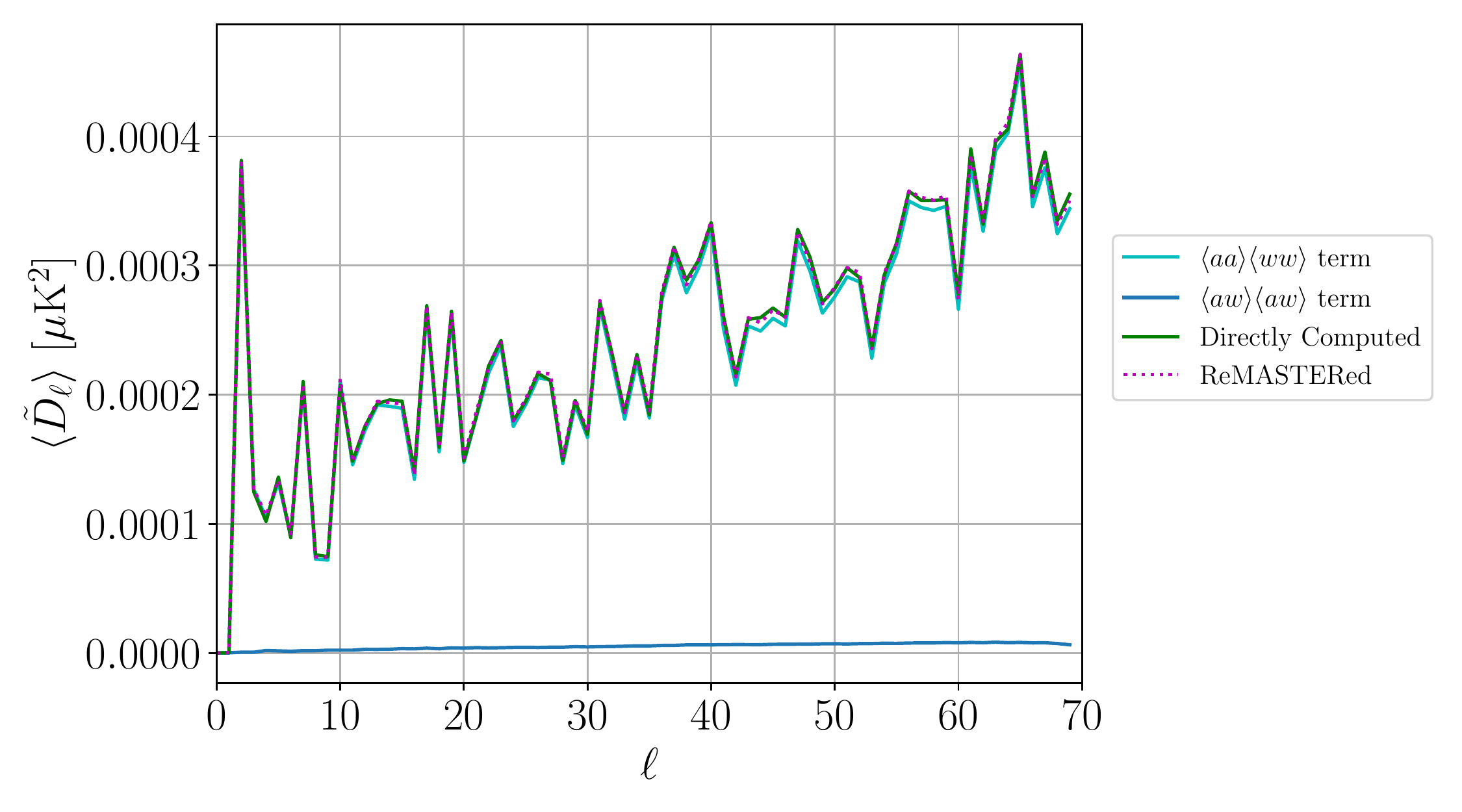}
    \caption{As in Fig.~\ref{fig:ISW_master} but for the ensemble average of 32 masked CMB maps, where for each realization the mask is equal to the CMB field plus a constant offset. This validates all the two-point functions in our result since the bispectrum and trispectrum are zero due to Gaussianity of both the CMB map and CMB + offset mask.}
    \label{fig:CMB_w_eq_aplusA}
\end{figure}

Another check is to let the mask simply be equal to the signal map plus some constant offset such that the mask is nonnegative. We keep this offset value fixed across different realizations. If the signal is a Gaussian random field, the bispectrum and trispectrum terms vanish and we can test the two-point terms alone:
\begin{equation}
    \langle \tilde{C}_{\ell_1}^{\rm GRF} \rangle = \frac{1}{4\pi} \sum_{\ell_2,\ell_3} (2\ell_2+1)(2\ell_3+1) \begin{pmatrix} \ell_1&\ell_2&\ell_3 \\ 0&0&0 \end{pmatrix}^2 \left[  \langle C_{\ell_2}^{aa}\rangle
    \langle C_{\ell_3}^{ww}\rangle +  \langle C_{\ell_2}^{aw}\rangle  \langle C_{\ell_3}^{aw}\rangle  \right]
\end{equation}

Using an initial WebSky CMB map (downgraded to $N_{\rm side}=64$) with \verb|healpy| to generate 32 random realizations and an offset for the mask of $4.56 \times 10^{-4}$, we obtain the result in Figure \ref{fig:CMB_w_eq_aplusA}.

\begin{figure}[t]
    \centering
    \includegraphics[width=0.80\textwidth]{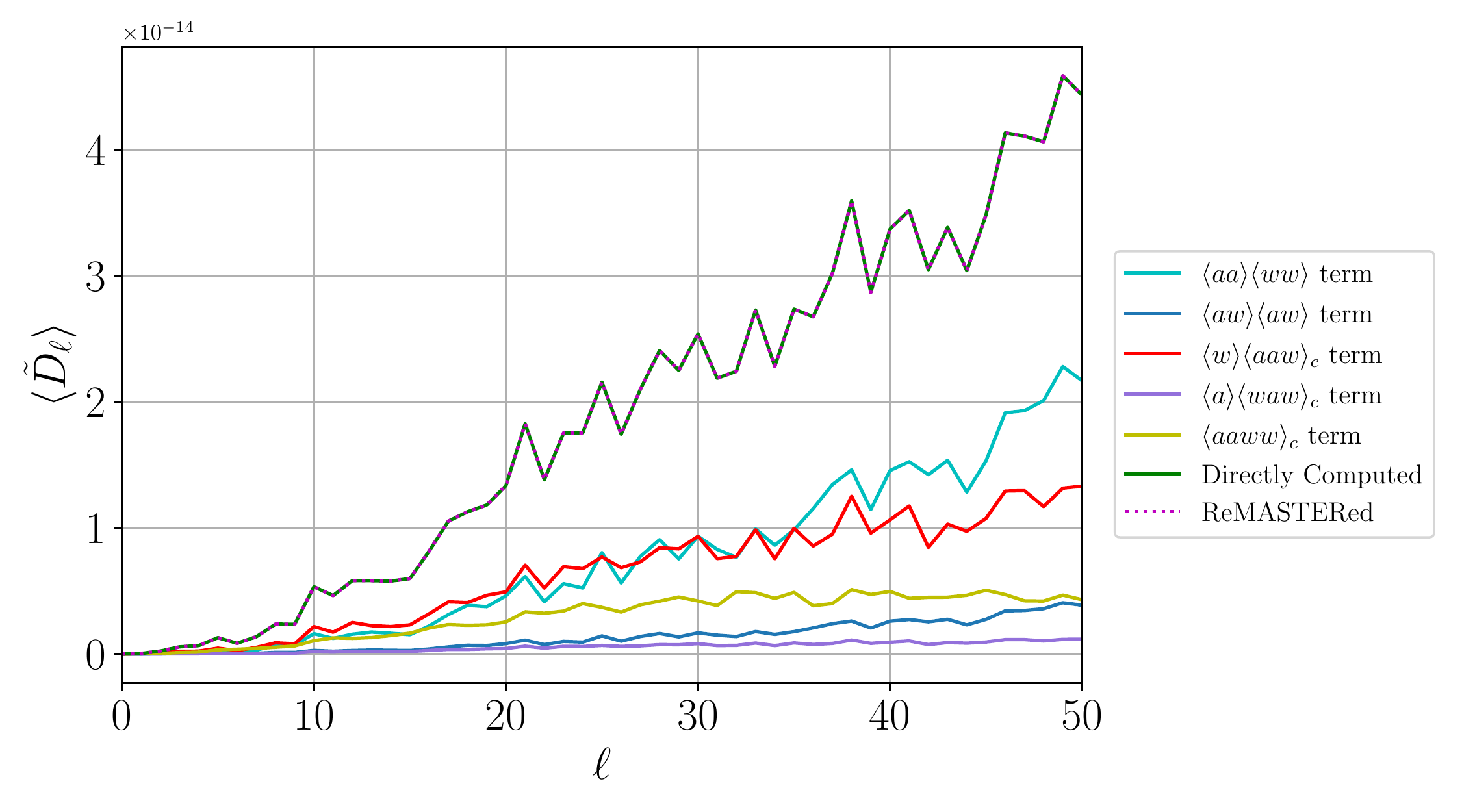}
    \caption{As Fig.~\ref{fig:ISW_master} but for a single masked tSZ effect map (given in dimensionless Compton-$y$ units), where the mask is equal to the tSZ field plus a constant offset. This example is illustrative of the potential importance of the bispectrum (solid red) and trispectrum (solid yellow) terms, which here have comparable magnitudes to the $\langle aa \rangle \langle ww \rangle$ term (solid cyan), the original MASTER result.}
    \label{fig:tSZ_w_eq_aplusA}
\end{figure}

As an extreme example, we can repeat the previous procedure but with a non-Gaussian field. In such a case, we would expect large contributions from the three-point and four-point terms. Previous analyses of biases due to correlated masks have also found that the tSZ field has especially high impact from non-Gaussian effects \cite{Lembo:2021}. Using a single tSZ effect map realization from \verb|halosky| (downgraded to $N_{\rm side}=128$) and a constant offset of $10^{-6}$ for the mask, we obtain the result in Figure \ref{fig:tSZ_w_eq_aplusA}. As before, we could have done this operation as an ensemble average (as one would do in practice), as \verb|halosky| generates several independent tSZ field realizations. We use the single-realization result only to demonstrate correctness.

%%%%%%%%%%%%%%%%%%%%%%%%%%%%%%%%%%%%%%%%%%%%%%%%%%%%%%%%%%%%
%%%%%%%%%%%%%%%%%%%%%%%%%%%%%%%%%%%%%%%%%%%%%%%%%%%%%%%%%%%%
\section{Biases on True Power Spectra from MASTER}
\label{sec.appendix_true_cl_ratio}

\begin{figure}[H]
    \centering
    \includegraphics[width=0.90\textwidth]{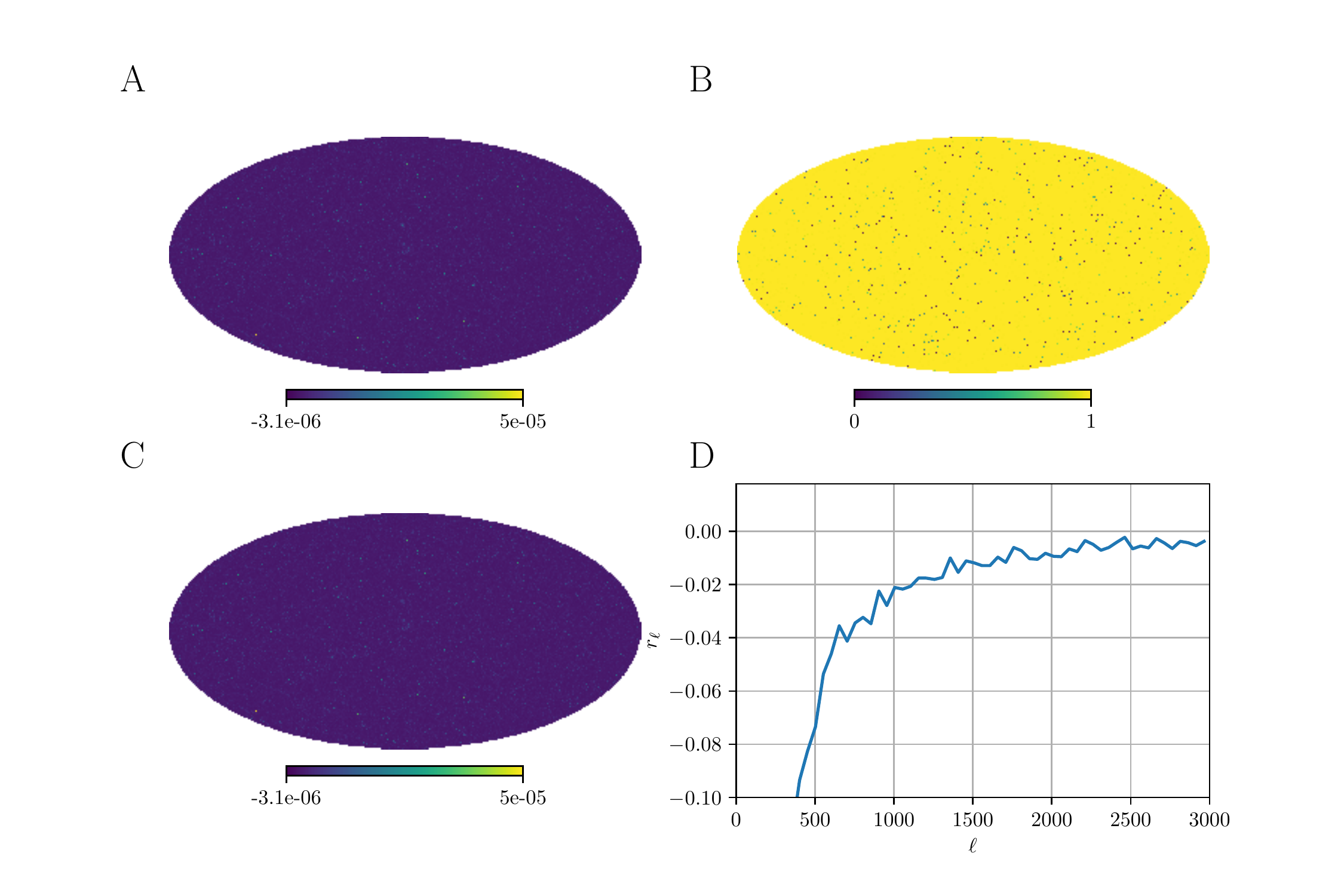}
    \caption{Same as Fig.~\ref{fig:tSZ_maskIR_maps}, but for the full WebSky Compton-$y$ map with resolution parameter $N_{\mathrm{side}}=4096$ and IR sources down to 5 mJy masked with 2 arcmin holes. The mask is apodized with a cosine apodization of width 2 arcmin. The correlation coefficient of this Compton-$y$ map and IR source mask is shown out to $\ell=3000$ with an $\ell$-space binning of $\Delta \ell = 50$. }
    \label{fig:high_res_tsz_mask_IR}
\end{figure}

In \S \ref{sec.isw_threshold} and \S \ref{sec.y_mask_ir} we examine ratios of the forward-modeled pseudo-$C_\ell$ to the directly computed pseudo-$C_\ell$ for both the MASTER and reMASTERed cases. As discussed in \S \ref{sec.discussion}, in actual data analyses that use MASTER, the mode-coupling matrix is inverted to obtain the ``true" $C_\ell$ from the observed pseudo-$C_\ell$. Here we examine biases on the true $C_\ell$ obtained via MASTER for the ISW field with the threshold mask and tSZ field with the IR source mask that were discussed previously. Since the tSZ effect is dominant at small scales relative to those we have considered thus far, to get a more realistic sense of biases that result from MASTER for this use case, here we use the full original WebSky Compton-$y$ map with resolution parameter $N_{\mathrm{side}}=4096$, masking IR sources with flux $\geq 5$ mJy with holes of radius 2 arcmin, thus masking a total of 67296 sources. The mask is then apodized with a cosine apodization of width 2 arcmin. This map, mask, masked map, and binned correlation coefficient through $\ell=3000$ (in bins of $\Delta \ell = 50$) are shown in Fig.~\ref{fig:high_res_tsz_mask_IR}. We note that the magnitude and sign of the correlation coefficient at high $\ell$ are dependent on the size of the holes used in the mask construction as well as the scale of the mask apodization.

\begin{figure}[H]
    \centering
    \includegraphics[width=0.60\textwidth]{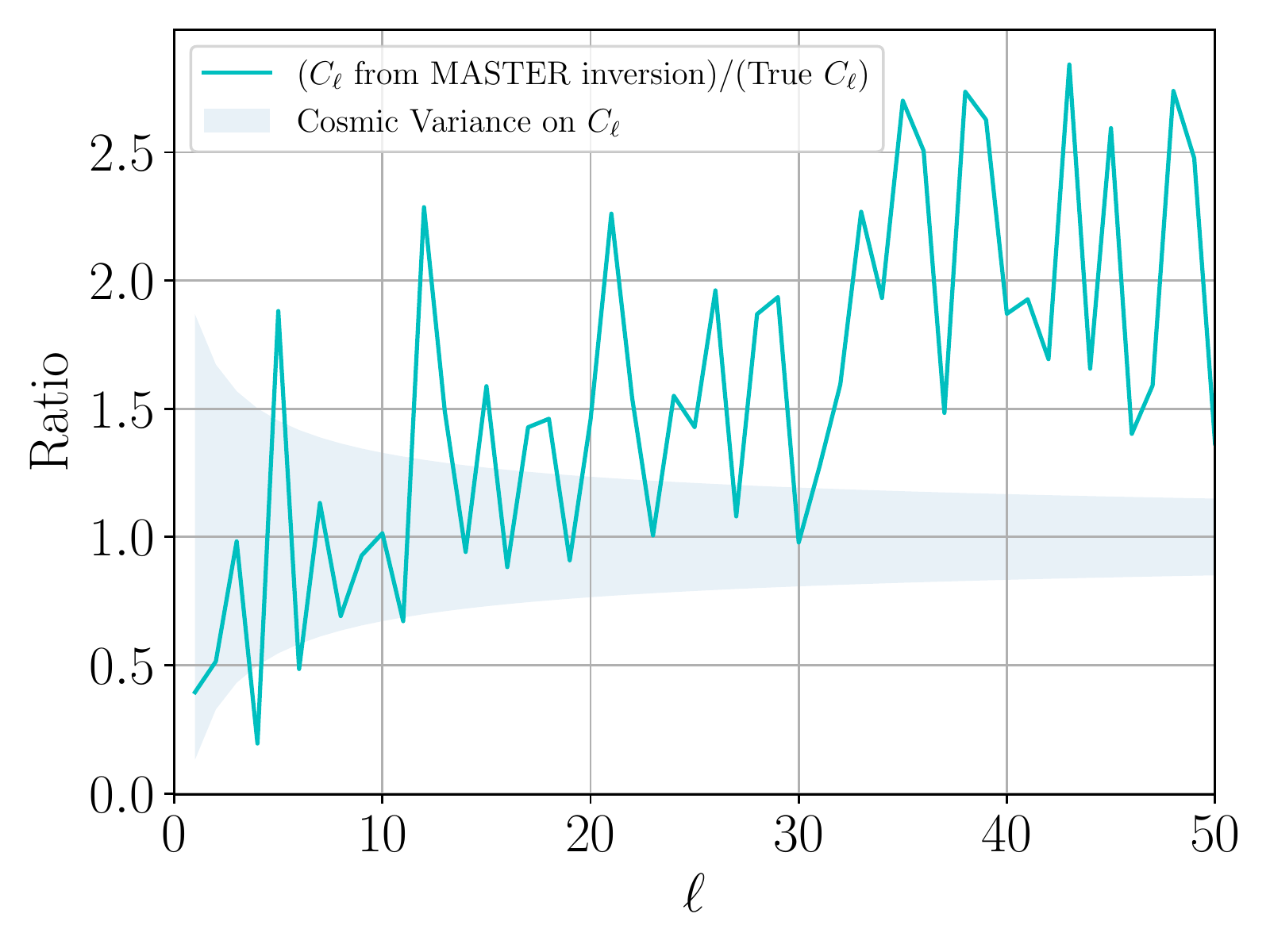}
    \caption{Ratio of $C_\ell$ obtained via inversion of the MASTER equation to the true $C_\ell$ for the ISW field with a threshold mask. The region between the bounds of error due to (the Gaussian approximation to) the cosmic variance of true $C_\ell$ is shaded in light blue.}
    \label{fig:ratio_true_cl_isw}
\end{figure}

The $C_\ell$ from MASTER are calculated by inverting the MASTER mode-coupling matrix and multiplying by the directly calculated pseudo-$C_\ell$. Fig.~\ref{fig:ratio_true_cl_isw} shows the ratio of $C_\ell$ obtained via MASTER to the true $C_\ell$ for the ISW field with a threshold mask. Letting $C_\ell$ be the true $C_\ell$, (the Gaussian approximation to) the cosmic variance is calculated as $\sigma^2_{C_\ell} = \frac{2}{(2\ell+1)f_{\rm sky}} C_\ell^2$, where $f_{\rm sky}$ is the unmasked sky fraction, and the bounds of the cosmic variance on the ratio plot are $(C_\ell \pm \sigma_{C_\ell})/(C_\ell)$.

\begin{figure}[H]
    \centering
    \includegraphics[width=0.60\textwidth]{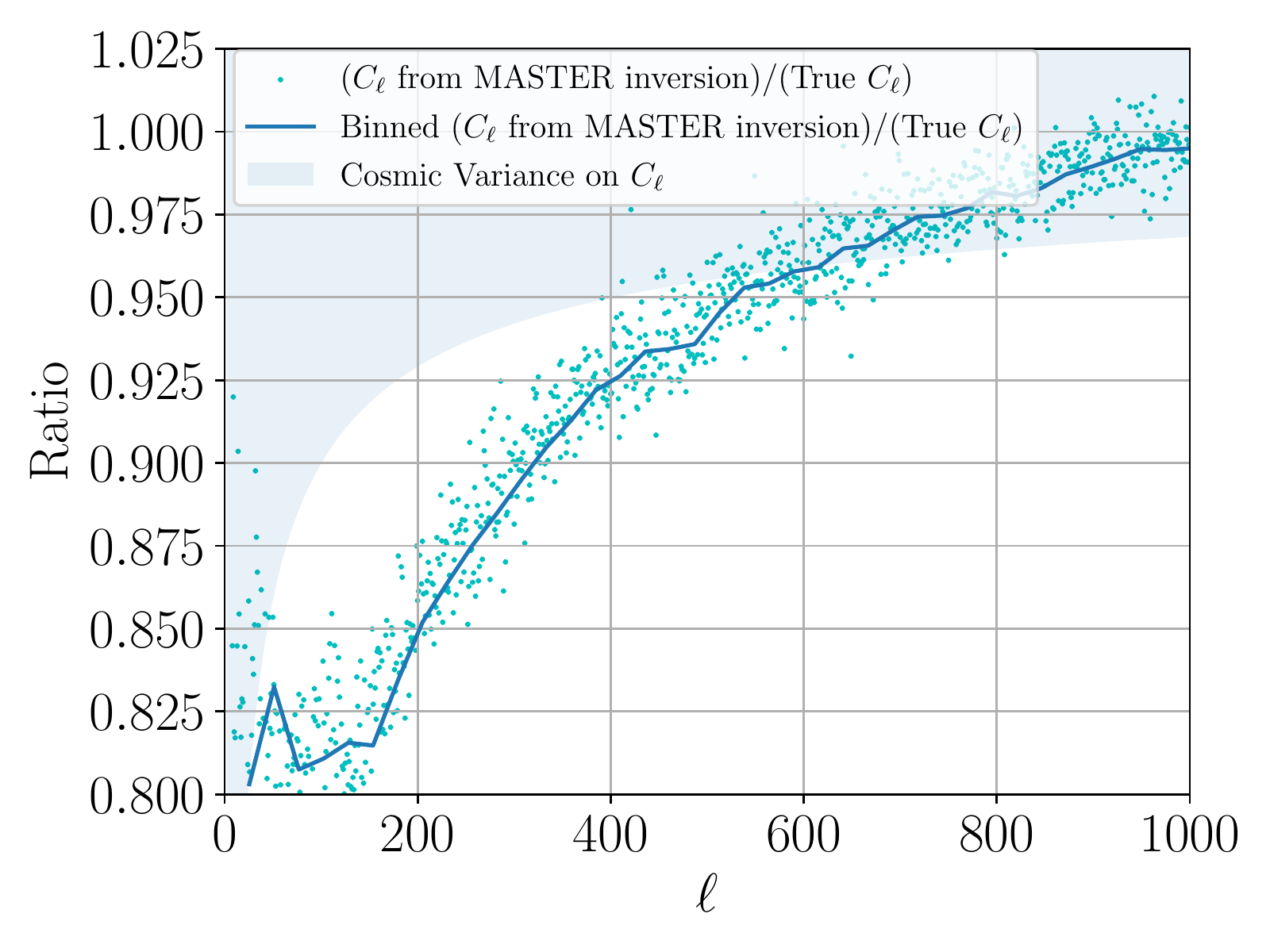}
    \caption{Same as Fig.~\ref{fig:ratio_true_cl_isw}, but for the Compton-$y$ field with an IR source mask. Here we use the WebSky Compton-$y$ map with resolution parameter $N_{\mathrm{side}}=4096$, masking IR sources with flux $\geq 5$ mJy with 2 arcmin holes. The ratios of $C_\ell$ obtained via inversion of the MASTER equation to the true $C_\ell$ at each multipole are shown as cyan points. The solid blue line shows these ratios in bins of width $\Delta \ell = 25$. }
    \label{fig:ratio_true_cl_tsz}
\end{figure}

Fig.~\ref{fig:ratio_true_cl_tsz} shows this ratio for the tSZ field with an IR source mask, with biases from the MASTER approach shown out to $\ell=1000$. We note the Gaussian-only variance is an underestimate at low $\ell$, where the connected trispectrum of the tSZ signal is large (\textit{e.g.},~\cite{Shaw2009,Hill:2013baa}). Full evaluation of this trispectrum is important for parameter inference from the tSZ power spectrum, but is not included in this plot for simplicity.

From both of these figures, it is clear that the MASTER result, as used in actual data analyses, can give large biases on the true $C_\ell$ for maps with correlated masks. This necessitates the use of our result to avoid such biases.

\end{appendices}
\bibliography{refs}
\end{document}